\documentstyle[epsfig,here]{elsart}
\begin{document}
\begin{frontmatter}
\title{The Optical Module of the Baikal \\
       Deep Underwater Neutrino Telescope}

\author[vier]{R.I. Bagduev},
\author[funf]{V. Balkanov},
\author[acht]{I.A. Belolaptikov},
\author[eins]{L.B. Bezrukov},
\author[funf]{N.M. Budnev},
\author[eins]{B.A. Borisovets},
\author[eins]{G.V. Domogatsky},
\author[eins]{L.A. Donskych},
\author[eins]{A.A. Doroshenko},
\author[eins]{A.A. Garus},
\author[drei]{A.V. Golikov},
\author[sechs]{B.M. Gluchovskoj$(\dag)$},
\author[zwei]{R. Heller},
\author[drei]{V.B. Kabikov},
\author[vier]{M.P. Khripunova},
\author[eins]{A.M. Klabukov},
\author[eins]{S.I. Klimushin},
\author[funf]{A.P. Koshechkin},
\author[drei]{L.A. Kuzmichov},
\author[vier]{G.V. Lisovski},
\author[eins]{B.K. Lubsandorzhiev},
\author[zwei]{T. Mikolajski},
\author[drei]{E.A. Osipova},
\author[eins]{P.G. Pokhil},
\author[funf]{P.A. Pokolev},
\author[vier]{P.A. Putilov},
\author[zwei]{Ch. Spiering\thanksref{talk}},
       \thanks[talk]{corresponding author: phone +49 33762 77218, e-mail
  csspier@ifh.de}
\author[vier]{Z.I. Stepanenko},
\author[zwei]{O. Streicher},
\author[zwei]{T. Thon},
\author[sieben]{A.A. Vorobiev},
\author[zwei]{R. Wischnewski}.

\address[eins]{Institute for Nuclear Research, Russian Acad. of
  Science (Moscow\
,Russia)}
\address[zwei]{DESY Zeuthen (Zeuthen,  Germany)}
\address[drei]{Moscow State University (Moscow, Russia)}
\address[vier]{KATOD Laboratory (Novosibirsk, Russia)}
\address[funf]{Irkutsk State University (Irkutsk, Russia)}
%$^6$ Tomsk Polytechnical Institute (Tomsk, Russia),\\
\address[sechs]{Moscow Electro Lamp Factory  MELS (Moscow, Russia)}
\address[sieben]{Russian State Chemico-Technological University
  (Moscow, Russia\
)}
\address[acht]{Joint Institute for Nuclear Research (Dubna, Russia)}

\begin{abstract}
A deep underwater Cherenkov telescope has been operating 
since 1993 in stages of growing size at 1.1 km depth in Lake Baikal. 
The key component of the telescope is the Optical Module (OM)
which houses the highly sensitive phototube {\it QUASAR-370}.
We describe design and parameters of the
{\it QUASAR-370}, the layout of the optical module,
the front-end electronics and the calibration procedures, and present
selected results from the five-year operation underwater.
Also, future developments with respect to a telescope
consisting from several thousand OMs are discussed.
\end{abstract}
\end{frontmatter}

\section{Introduction}
The Baikal Neutrino Telescope is being deployed in the
Siberian Lake Baikal, about 3.6 km from shore at
a depth of 1.1 km \cite{APP}, \cite{Proposal}. The central 
mission of the project is
the detection of extraterrestrial sources of 
high energy neutrinos. 
Other fields of interest \cite{Physics} are the search for 
neutrinos from WIMP annihilation in the Earth or the Sun,
for neutrino oscillations, and for slowly moving bright objects like
GUT monopoles. Standard cosmic ray physics with muons
generated in the atmosphere is covered as well as 
limnological and ecological questions.

In deep underwater detectors, clear water  serves
as target material for neutrino interactions,  as Cherenkov
radiator for charged particles, and as a shield against
atmospheric muons and sunlight. Energetic neutrinos are 
detected easiest by mapping the  
Cherenkov light from muons produced in charged current 
interactions. "Mapping" means measurement of the
photon arrival times at photodetectors distributed over a
large volume. The feebleness of the
light signal requires a large-area, large-acceptance
light detector with single photoelectron resolution. 
Mapping of the Cherenkov cone with a spatial 
accuracy not worse than the OM diameter requires
a time resolution of a few nano\-seconds. The water depth
demands pressure protection of the sensor.

The present paper describes design and operation
of the components of the optical module (OM) 
most of which have been
developed within our collaboration. After a short
presentation of the telescope in section 2,
section 3 covers the design and the parameters
of the phototube {\it QUASAR-370}. Section 4 gives the construction
of the OM, section 5 describes the electronics and  
discusses the operational principle of two PMTs switched
in coincidence. Section 6 presents results from the
different methods of OM calibration, whereas in section 7 
the long-term operation underwater is evaluated and
some selected results of the telescope operation are given. 
Section 8 summarizes the results and sketches routes of
further development.

\section{The Telescope {\it NT-200}}
After numerous experiments with prototype configurations
\cite{Proposal}, in
April 1993 we deployed a first underwater detector allowing
three-dimensional track reconstruction of muons. This array {\it NT-36}
consisted of 36 OMs at 3 strings \cite{NT-36}. 
It was replaced in 1994 by a slightly modified version, in
1995 by a 72-OM array, in 1996 by {\it NT-96} consisting of
96 OMs at 4 strings, and in 1997 by a
144-OM array.  These detectors have been
steps towards the Neutrino Telescope {\it NT-200}
\cite{APP,Proposal} with a total of 192 OMs. {\it NT-200}
was completed in April 1998 and is presently taking data.
It is sketched in Fig. 1.

The OMs consist of a pressure glas housing equipped with the {\it
  QUASAR-370}.
They are grouped in pairs along the strings.
The two PMTs of a pair are switched in coincidence, defining a {\it
  channel}.
The "constructional" basic building block (called {\it "svjaska"})
consists of two pairs of OMs,
and the svjaska electronics module, {\it SEM}, which houses
control units, power supply and the front-end electronics.

\begin{figure}[H]
\centering
  \mbox{\epsfig{file=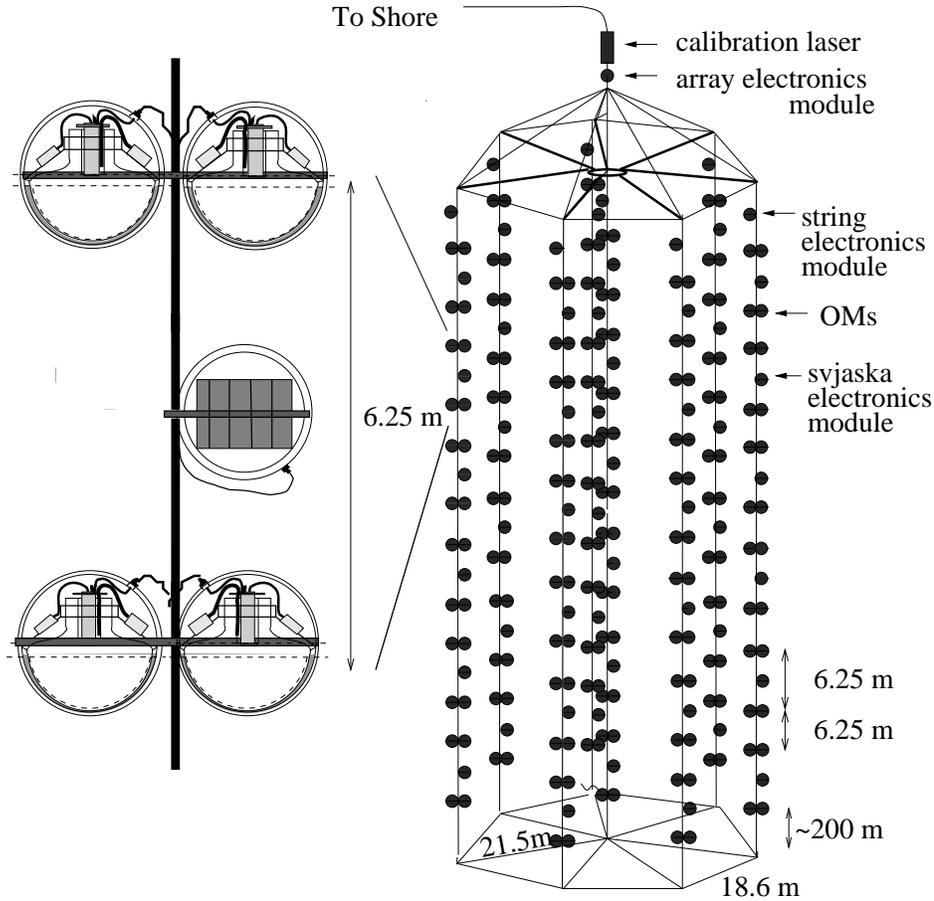,height=12.5cm, angle=-90}}
\caption[1]{\small
Schematic view of the Baikal Telescope {\it NT-200}.
The expansion lefthand shows 2 pairs of OMs 
("svjaska"), with the svjaska electronics module
housing parts of the control and read-out electronics.
}
\end{figure}

\bigskip

A {\it muon trigger} is formed if $\geq m$ channels are hit
within a time window of 500 nsec (this is about twice
the time a relativistic particle needs to cross the {\it NT-200}
array). The  value $m$  is typically set to 3 or 4. Times 
and amplitudes are digitized in the string electronic modules.

A second system {\it "monopole trigger"}
searches for counting rate patterns characteristic for slowly moving
bright particles like nuclearities or GUT magnetic monopoles
catalyzing proton decays. Depending on the velocity of the
object, such events could cause enhanced counting rates
in individual channels during time intervals of 0.1\,-\,0.8 msec,
separable from Poissonian noise.

\section{The Phototube}

\subsection{Construction and operational principle}
The {\it QUASAR-370} 
consists of an electro-optical preamplifier
followed by a conventional photomultiplier (PMT) - see Fig. 2.
In this hybrid scheme, photoelectrons from a large hemispherical cathode
with  $>$ 2$\pi$ viewing angle are accelerated by 25 kV to a fast, high
gain scintillator which is placed near the center of the glass
bulb. The light from the scintillator is read out by a small
conventional PMT named {\it UGON}. 
One photoelectron emerging from the hemispherical
photocathode yields typically 25 photoelectrons in the conventional
PMT. This high multiplication factor of the electro-optical preamplifier
results in an excellent single electron resolution -- important
for the detection of low level light pulses and background
suppression. Due to the fast acceleration of primary photoelectrons
by 25 kV high voltage, the time jitter can be kept low. This is
most important for accurate track reconstruction. Last not
least, the tube is almost insensitive to the
Earths magnetic field.

%\begin{figure}[H]
\begin{figure}[b]
\centering
  \mbox{\epsfig{file=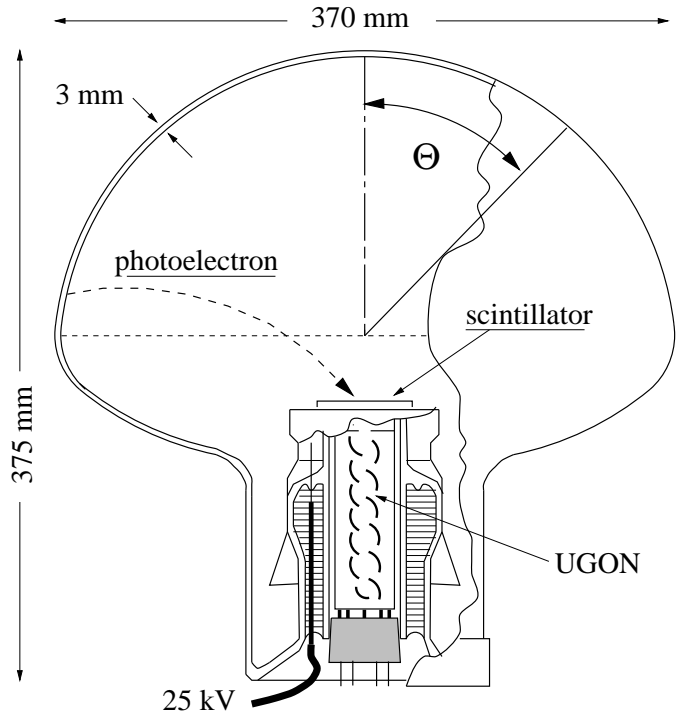,height=8.8cm, angle=-90}}
\caption[4]{\small
Cross section of the QUASAR-370 tube
}
\end{figure}

A hybrid phototube of this kind, XP2600, was first
developed by PHILIPS \cite{XP-1,XP-2}. After first experience
with the XP2600 we followed their  basic  design 
and developed  the {\it "QUASAR"}.

First versions of the {\it QUASAR}-tube had a spherical shape, with
diameters of 30\,cm ({\it QUASAR-300}) and 35\,cm 
({\it QUASAR-350}), respectively.
The latest version  -- {\it QUASAR-370}  --  has
a nonspherical (mushroom) shape of  the glass  bulb  to  provide  more
isochronic photoelectron trajectories. 
%Fig. 4 shows the single
%photoelectron transit time distributions for point-like illumination
%of the photocathode at different zenith angles, 
Fig. 3 gives  the
measured relative transit time as a function of the zenith
angle. The transit time differences are minimized to
$\le$ 2.0 nsec.

%\vspace{0.8cm}
\begin{figure}[H]
\begin{minipage}[b]{6.7cm}
\epsfig{file=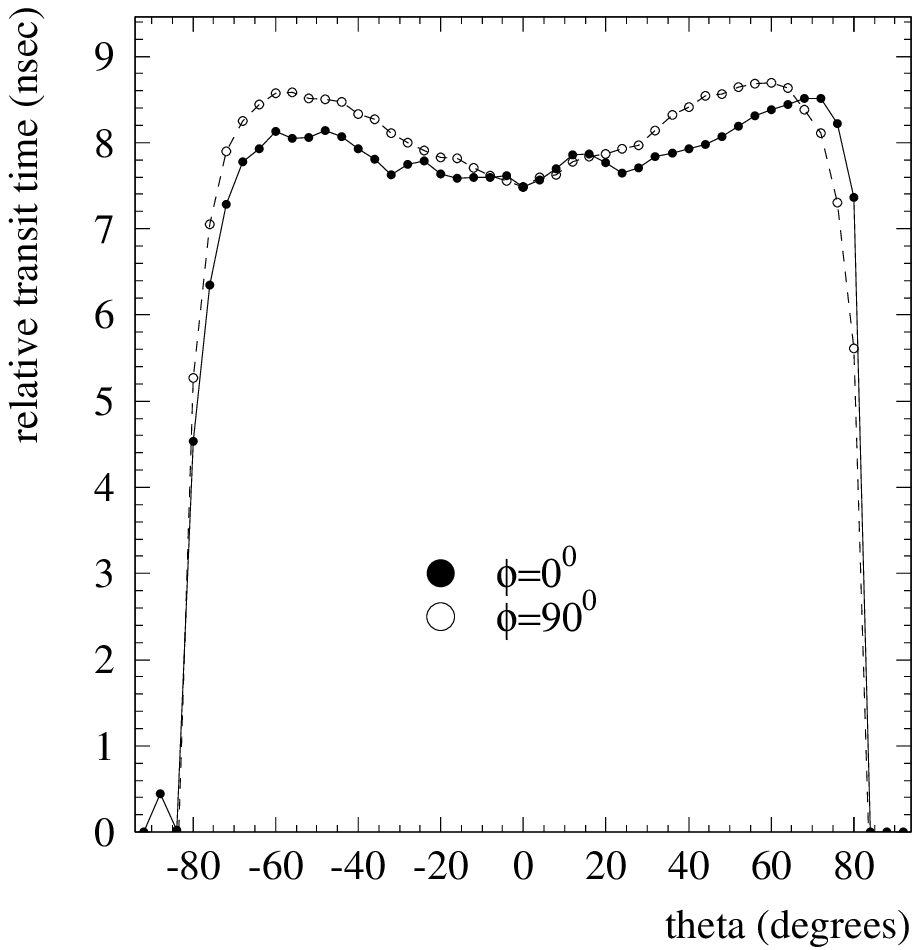,width=6.5cm}
 \caption[10]{Measured cathode transit time difference for 
  {\it QUASAR-370} (N$^o$ 254) versus zenith angle $\theta$, for 
   two azimuth angles $\Phi$.}
\end{minipage}
\hfill
\begin{minipage}[b]{6.7cm}
\epsfig{file=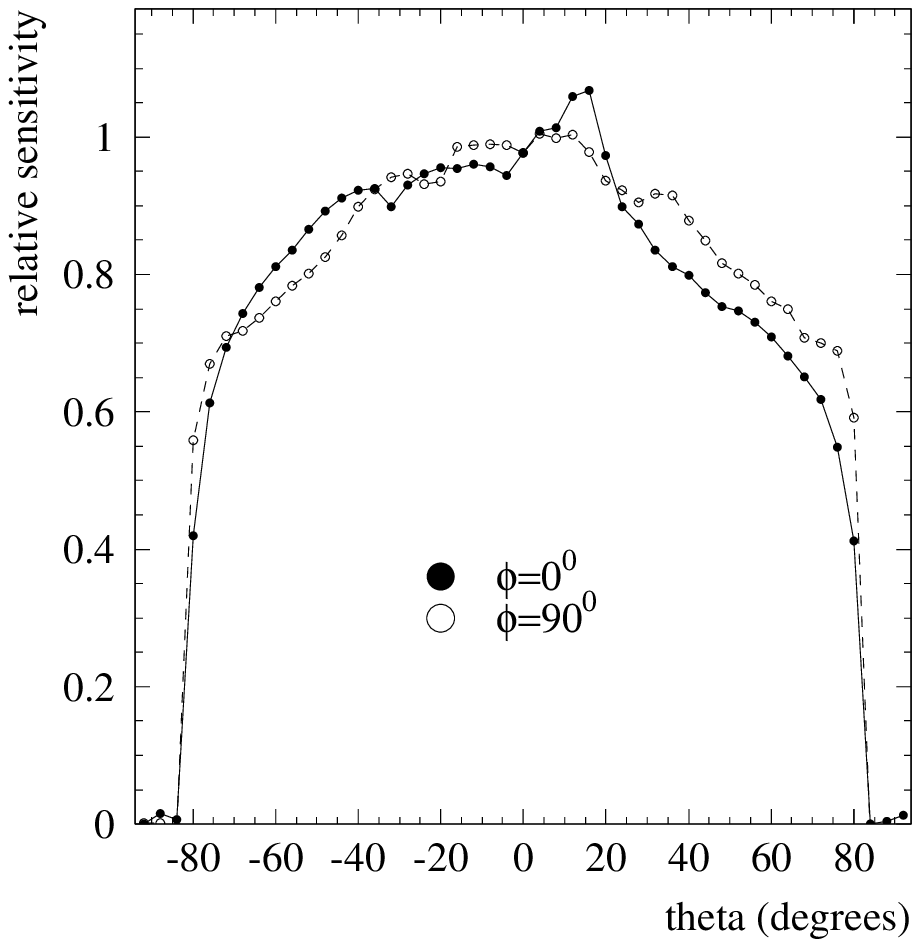,width=6.5cm}

\vspace*{0.5cm}
\caption [11]{Relative sensitivity of {\it QUASAR-370} (N$^o$ 254)
            vs. zenith angle $\theta$, for two azimuth angles $\Phi$.}
\end{minipage}
\end{figure}

The spherical "face" region of the bulb has a diameter of 37 cm. 
Modifications towards the mushroom form are made at large zenith
angles.
The bulb is manufatured from  borosilicate  glass  S49-1
by  the EKRAN  company,  Novosibirsk.  The photocathode
is   of the bialkali  type (\mbox{K$_{2}$CsSb}). Its  spectral
response  is  typical  for  this  type   of
photocathode,  with a  maximum  at  $\lambda$ = 400\,-\,420 nm.  The
spectral sensitivity exceeds  60  mA/W  at  $\lambda=420$~nm
which corresponds to  $\sim20\,\%$  quantum  efficiency.  The 
non-uniformity of the response across the photocathode  is  less  than
30\,\% (see Fig. 4).

The luminescent screen is made from pulverized  phosphor,
Y$_{2}$SiO$_{5}$(YSO). This scintillator
has a light yield of 20\,-\,30\,\%  relative to NaI(Tl) and 30\,-\,40 ns decay
time. 

\subsection{Single Photoelectron Resolution}
The single photoelectron resolution of the {\it QUASAR-370} is defined
mainly by the gain $G$ of the electro-optical preamplifier:

\begin{equation}
G=\frac{\mbox{number of photoelectrons detected  by  small
         PMT}}{\mbox{number of  photoelectrons  at  the 
 preamplifier
         photocathode}}.
\label{eq:1}
\end{equation}

%\smallskip
%\noindent
%$G$ can be expressed as
%\begin{equation}
%G=Y(E_{e})\cdot\xi\cdot\eta^{\prime},
%\label{eq:2}
%\end{equation}
%with Y(E$_{e}$)   --   the
%number  of  photons generated in  the luminescent  screen    by   one
% photoelectron,   $\xi$   --    the photon    collection
% efficiency,
%and $\eta^{\prime}$  --  the effective  quantum  efficiency  of
%         the photocathode of the small  PMT
%($\eta^{\prime}=\eta\cdot\kappa\cdot\kappa^{\prime}$;
% $\eta$
% --  quantum  efficiency;  $\kappa$  --  collection  efficiency  of
%photoelectrons at first dynode; $\kappa^{\prime}$ --  photoelectron
%detection  efficiency by dynode system).

Figs. 5 and 6 show typical charge distributions for single-  and 
multi-photo\-elec\-tron pulses of a {\it QUASAR-370}. The high amplification
factor $G$ allows to separate pulses of one and two
photoelectrons and to identify even the shoulder from 3 p.e. events.
The light pulse has been generated by a light emitting diode.
The
distribution labeled "single p.e." has been obtained by attenuating
the LED to a level when only every tenth LED pulse triggered the {\it QUASAR}.

Averaged over 100 tubes, the mean values for
single photoelectron  resolution  {\it SPR},
peak-to-valley ratio {\it P/V},
and gain $G$ are

\begin{itemize}

\item {\it  SPR} $\approx$ 70\,\%  (FWHM), 
\item {\it P/V} $\approx$ 2.5,
\item {\it G} $\approx$ 25.

\end{itemize}

\bigskip

\begin{figure}[H]

\begin{minipage}[t]{6.7cm}
\epsfig{file=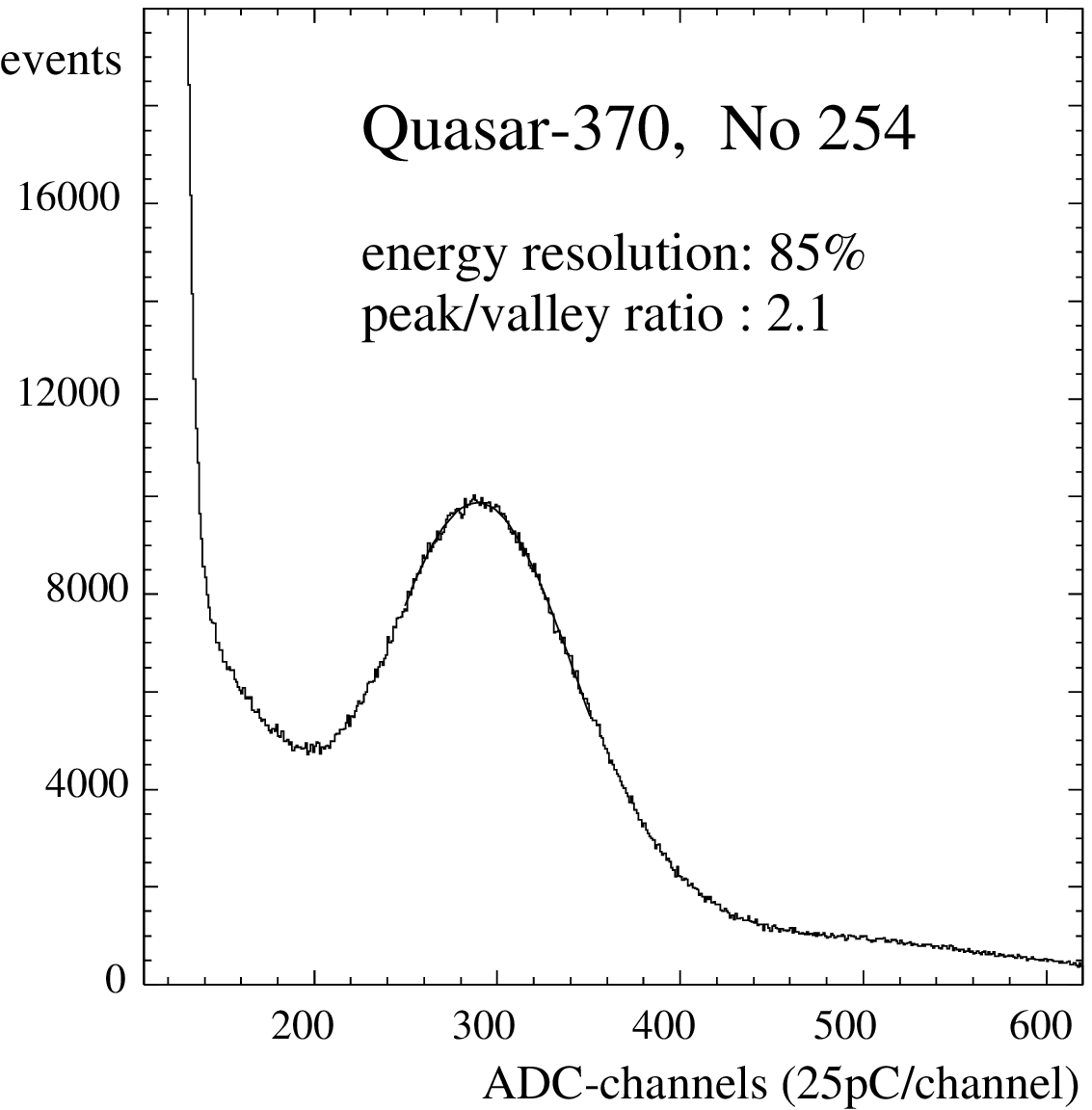,width=6.5cm}
 \caption [10]
 {Charge distribution for single p.e. events.}
\end{minipage}
\hfill
\begin{minipage}[t]{6.7cm}
\epsfig{file=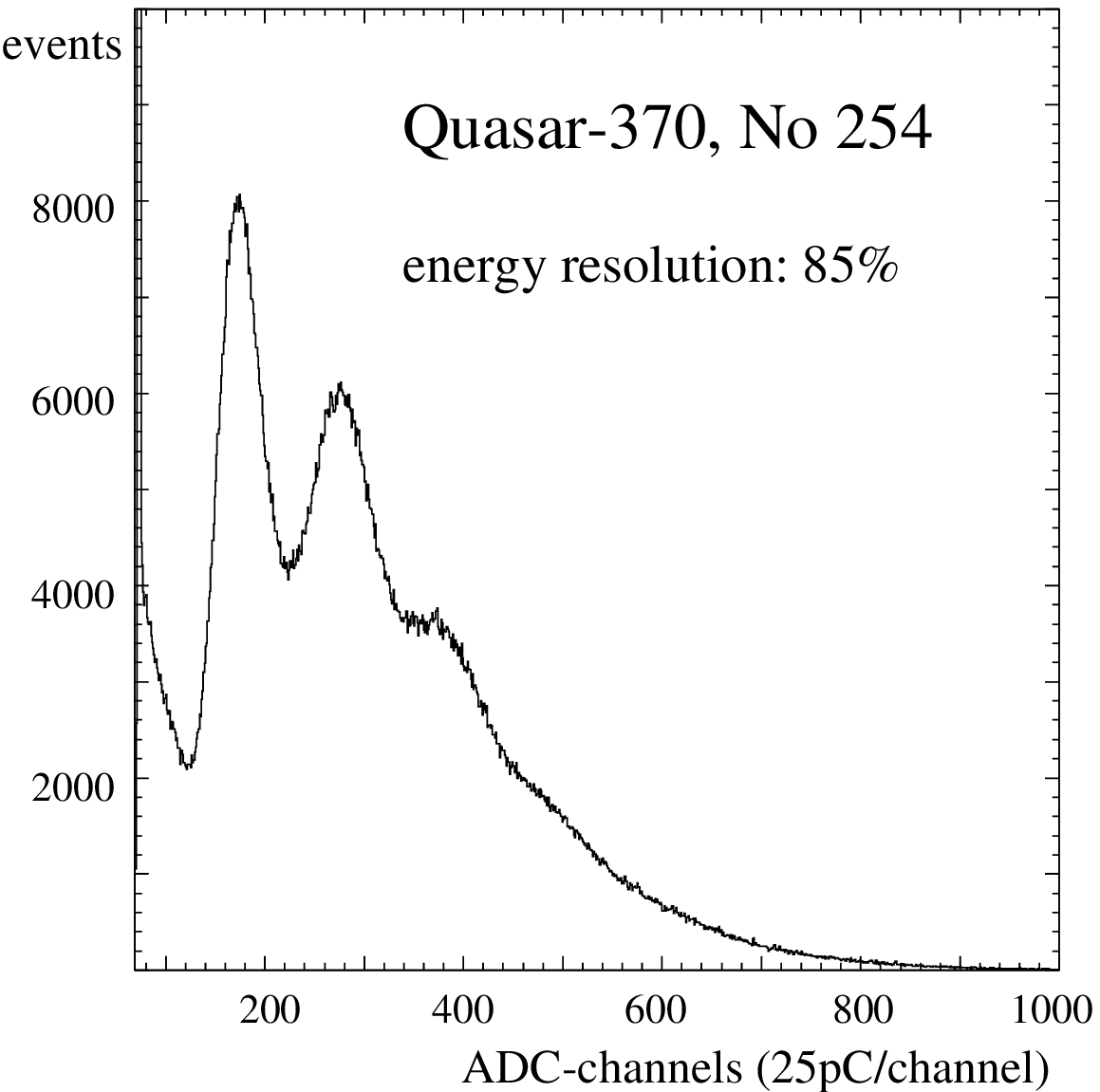,width=6.5cm}
\caption [11]
           {
Charge distribution for multi p.e. events.
}
\end{minipage}
\end{figure}

\subsection{Time Response}
A single photoelectron pulse of the {\it QUASAR-370}
 is a  superposition of  $G$
 single  photoelectron pulses of the small
 tube {\it UGON}, distributed exponentially in time:

\begin{equation}
P(t)=\frac{1}{\tau}\exp(-\frac{t}{\tau}),
\label{eq:3}
\end{equation}

with $\tau$ being the time constant of the scintillator.
Figs. 7 and 8 show the corresponding typical pulseforms 
of single and multi-photoelectron pulses.

\vspace*{0.5cm}
\begin{figure}[H]
\begin{minipage}[t]{6.7cm}
\epsfig{file=pmta.epsi,width=6.3cm}
 \caption [10]
  {
Typical pulse  form of a single p.e. pulse (10 mV/div vertically and
25 nsec/div horizontally).
}
\end{minipage}
\hfill
\begin{minipage}[t]{6.7cm}
\epsfig{file=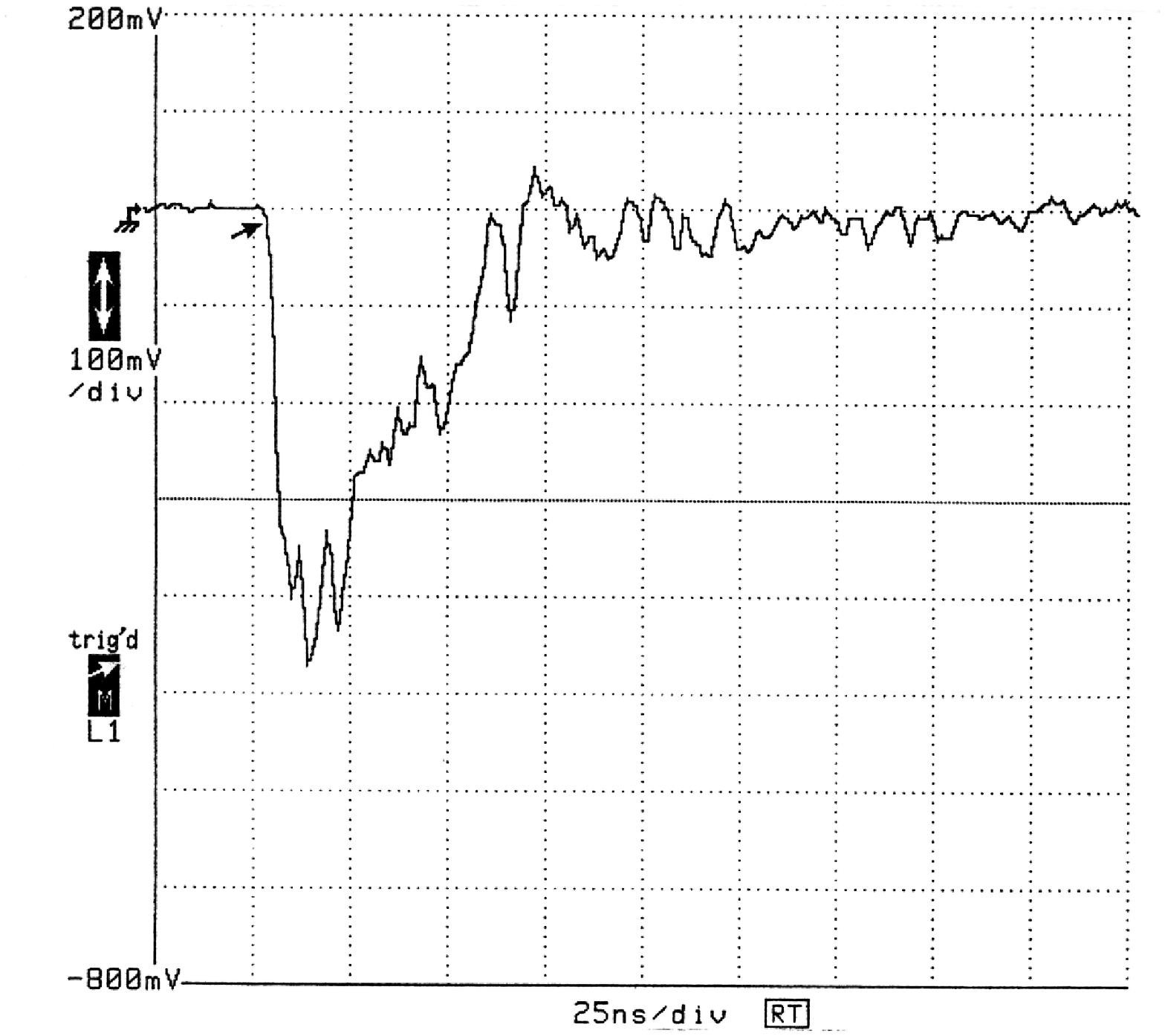,width=6.5cm}
  \caption [11]
  {
Typical pulse  form of a multi-p.e. pulse (100 mV/div vertically and
25 nsec/div horizontally).
}
\end{minipage}
\end{figure}

The  best  single  photoelectron  time  resolution  is
obtained by using a double threshold discriminator as sketched in 
Fig. 9. It consists of two discriminators with different thresholds 
and integration constants: a {\it timing} discriminator with a 
threshold of $0.25\,q_{1}$
and a {\it strobe} discriminator
with a threshold of $0.3\,Q_{1}$
($q_1$ and $Q_1$ are the most probable charges of a single
photoelectron pulse from the small PMT and from the big photocathode,
respectively). 

%\begin{figure}[H]
\begin{figure}[t]
\centering
  \mbox{\epsfig{file=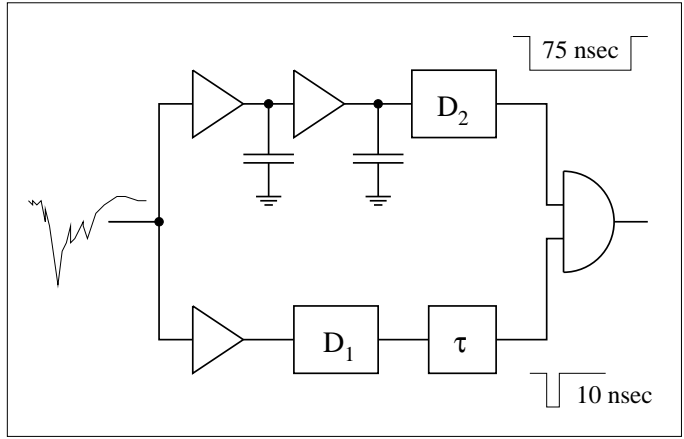,height=9cm, angle=-90}}
\caption[4]{\small
 Block diagram of the discriminator. D$_1$ -- timing
             discriminator with threshold 0.25 $q_1$, D$_2$ -- strobe
             discriminator with threshold 0.3 $Q_1$ (see text).
}
\end{figure}

The time is defined by the front of the  first  of the  $G$
single photoelectron pulses of the small PMT.
In this case, the transit time distribution for single photoelectron
 pulses with respect to the big photocathode  is described
     by 

\begin{equation}
W_{1}(t)=\frac{G}{\tau}\exp(-\frac{G}{\tau}t).
\label{eq:4}
\end{equation}

$W_1(t)$ is determined
by  the  scintillator  decay  time  constant  $\tau$  and  by the gain
$G$ of the electro-optical
preamplifier. For the best  tubes and an accelerating
voltage of 25 kV, $G$  is  about
 50, and the FWHM of $W_{1}(t)$ is 1.8 nsec for
point illumination. For typical tubes the transit time FWHM
is between 2 and 3.5 nsec. Fig. 10 shows the single photoelectron
transit time distribution for head-on full-cathode illumination.
The measured FWHM (3.8 nsec) is a convolution of the 
jitter for point illumination and the transit time differences from 
different parts of the  photocathode.

\vspace*{0.5cm}
%\begin{figure}[H]
\begin{figure}[b]
\centering
  \mbox{\epsfig{file=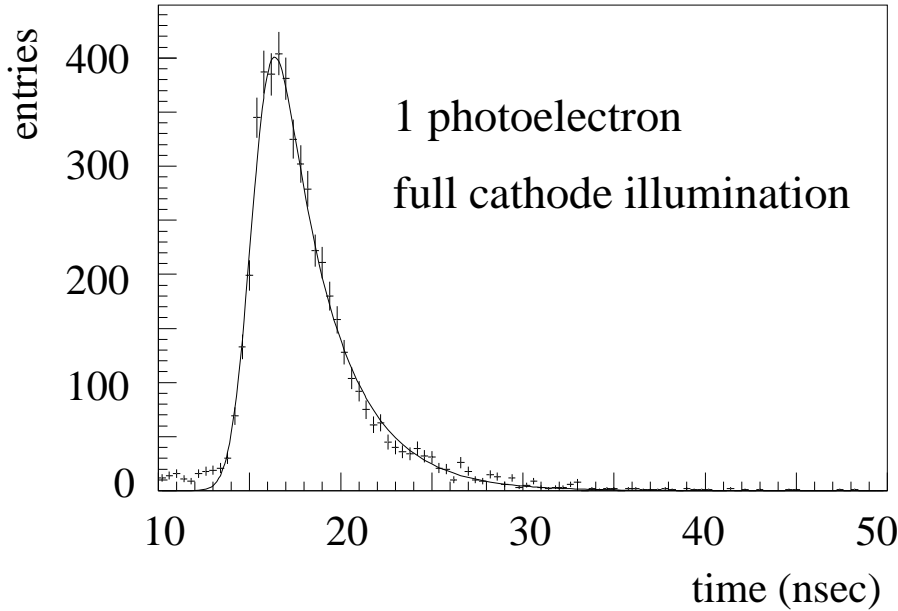,height=8cm}}
\caption[4]{\small
Single photoelectron transit time distribution for head-on,
full-cathode illumination of the photocathode of {\it QUASAR-370}
(N$^o$ 1). The FWHM is 3.8 nsec.
}
\end{figure}

We should note here  that  some  "late"  events
contribute to the tail  of the $W_{1}(t)$  distribution.
These events are due to backscattering of photoelectrons in
the luminescent screen. Elastically (or nearly
elastically) scattered electrons may leave the screen without
yielding a signal above the discriminator threshold. They  are bent
back by the electrical field in the electro-optical preamplifier
and hit the screen a second time. Due to the high voltage 
(25 kV), the  scale  of  delay  times
of late events in the {\it QUASAR-370} is considerably smaller than
in conventional PMTs -- about 10 nsec compared to
 30\,-\,100 nsec. 

The level  of
ordinary afterpulses in the {\it QUASAR-370} is substantially less\linebreak
($\le$ 2\%)  than
in conventional PMTs. The reasons are {\it a)}
the complete  vacuum  separation
 between the electro-optical preamplifier and the small  PMT,
and {\it b)} the low sensitivity of the photocathode to
backscattered X-ray photons of typically 10 keV (compared to
some 100 eV in conventional PMTs).  

Table I summarizes the main parameters of the {\it QUASAR-370}
and of the small PMT {\it UGON}.

\vspace*{0.5cm}
\begin{center}
{\bf \large Table I} \\ [0.5cm]
\begin{tabular}{|l|c|c|}
\hline
& QUASAR-370 & UGON \\
\hline
bulb material & borosilicate glass & borosilicate glass \\
photocathode & K$_2$CsSb & K$_2$CsSb \\
photocatode diameter & 37 cm & 2.5 cm \\
spectral sensitivity ant $\lambda$ = 410 nm & 60 mA/W & 60 mA/W \\
number of stages & 1 & 12/13 \\
gain & 25 & 10$^7$ \\
1-PE resolution & 70 \%  & --   \\
peak-to-valley ration (1PE)  & 2.5  & 1.3 \\
TT difference (center-edge) & $\le$1.5 nsec & $\le$ 1 nsec \\
TT jitter for 1 PE point illumination & 2 nsec & 2.2 nsec \\
noise rate ($\ge$ 0.25 PE, 20$^o$C) & 30 kHz & $\le$ 1 kHz \\
\hline
\end{tabular}
\end{center}

\section{Design of the Optical Module}

\subsection{General Description}
The OM basically consists of the {\it QUASAR-370} enclosed
in a transparent, nearly spherical pressure housing, see Fig. 11.
The optical contact between the photocathode region of the tube and
the pressure sphere is made by liquid glyzerine sealed with a layer of
polyurethane.
Apart from the PMT, the OM contains two HV supplies (25 kV and 2 kV)
for the hybrid PMT, a voltage divider, two preamplifiers, a calibration
LED and a vacuum probe. Each OM is electrically connected to the
Svjaska Electronics Module (SEM, see figs.1 and 15) by four electrical 
lines. They pass the signal
driving the LED from the SEM to the OM, and the PMT anode and dynode
signal from the OM to the SEM. The fourth cable supplies the low
voltages for the PMT HV-system and the preamplifier. A vacuum valve
(not shown in Fig. 11) allows to evacuate the sphere to 0.7 atm (see
section 4.2). 
The OM is fixed to the string by a steel frame locked via a shackle.
Fig. 12 shows a photography of an OM pair.

\vspace*{0.5cm}
\begin{figure}[H]
\centering
  \mbox{\epsfig{file=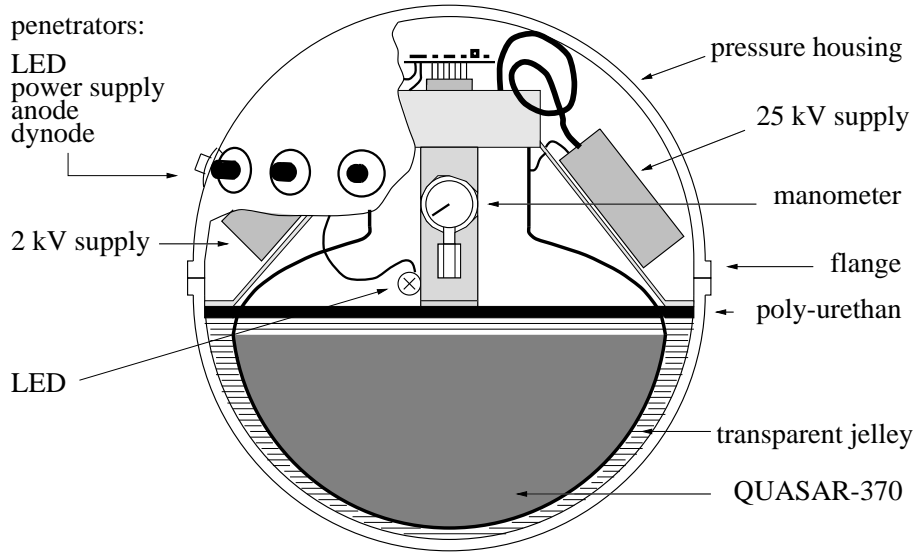,height=12cm, angle=-90}}
\caption[4]{\small
Schematical view of an Optical Module
}
\end{figure}

\subsection{Pressure Housing}
{\it Early approaches}

For the single string installations operated up to 1989 at Lake
Baikal, cylindrical housings made from epoxy reinforced fiber glass
have been used. These OMs
housed two 15-cm tubes with flat window, facing
upward and downward, respectively. The PMTs were covered with
end caps made from plexiglas. Limits on the flux of GUT monopoles
catalyzing baryon decay as well as a variety of limnologically
relevant results have been obtained with single strings carrying
these OMs \cite{Proposal}. With the advent of big 
hemispherical tubes this solution was discarded.
In parallel to the tests of the early versions of 
the {\it QUASAR} \cite{Q-300}, we considered a
pressure resistant phototube
%glass sphere tailored to  combine the functions of {\it i)} 
%a pressure housing and {\it ii)} the bulb of the electro-optical 
%preamplifier of the hybrid phototube.
and tested pilot samples
of appropriate glass spheres with 37 cm diameter and 0.8\,-\,1 cm
wall thickness \cite{Dor93}. However, in order to have more
flexibility for future improvements of the phototube,
we soon decided  to use separate PMTs and pressure housings. 

\begin{figure}[H]
\centering
  \mbox{\epsfig{file=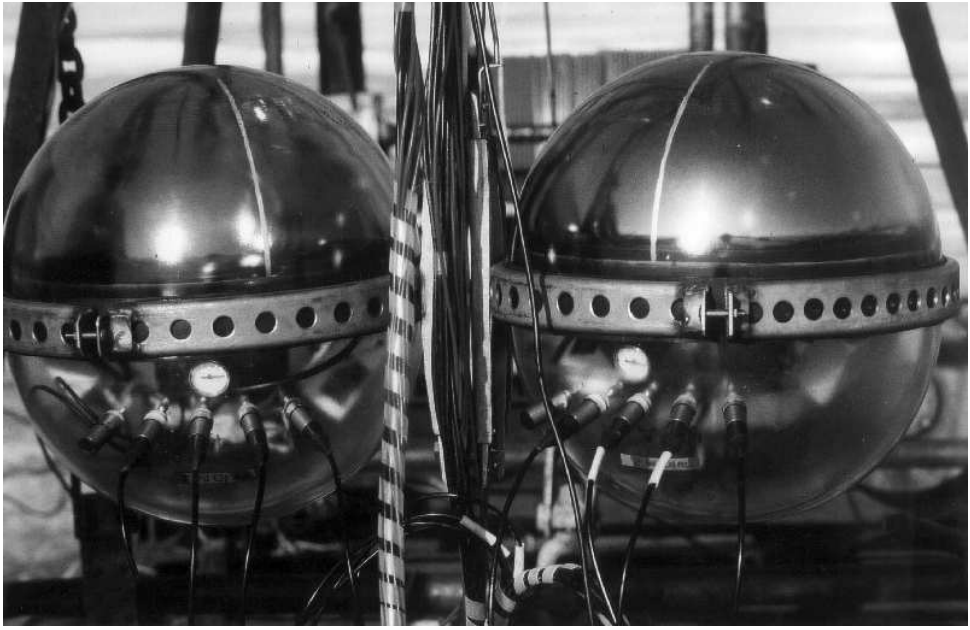,height=8.3cm}}
\caption[1]{\small
A pair of Optical Modules of the Baikal telescope.
The photocathodes point
upward. Cleary seen are the four electrical feedthroughs
and the vacuum valve lefthand.
}
\end{figure}

{\it Present design}

Traditional housings for large deep underwater
instruments consist of two hemispheres whose
equatorial surfaces are carefully ground to match each other.
$15^{\prime \prime}$ spheres are 
produced by BENTHOS Inc, USA, and
Jena-Glass, Germany (VITROVEX).
In 1987, together with the EKRAN company (Novosibirsk)
we started the design of an own pressure housing.
It consists from the same S49-1 borosilicate glass used
for the bulb of the {\it QUASAR-370}.
Its refractive index is 1.47\,-\,1.48. 
Since we developed the housings for our own purpose, we
could optimize form and dimensions to fit 
the demands of the Baikal experiment. The originally
spherical form was elongated by adding a cylindrical
part of 2 cm to the equator of each hemisphere. This 
allowed {\it i)} to avoid space problems when mounting the
tube with its high voltage module into the housing and
{\it ii)} additionally to use the same housing for the
underwater modules housing electronics crates.
The elongated housing is superior to a sphere with bigger
diameter since the layer of immersion material
between tube bulb and pressure housing can be kept thinner.
This as well as the small wall thickness (1.0\,cm compared to
1.4\,cm for  BENTHOS and VITROVEX)
results in a low light absorbtion. The 1.4\,cm spheres withstand 
a water depth of 6.7\,km, the wall thickness of the EKRAN sphere
is sufficient to work at all depths in Lake Baikal (max. 1632\,m).
The transmission at 500\,nm is 87 \% for the EKRAN sphere,
and 83 \% for the other two spheres.

In order to simplify the construction of the metallic belt
used to clamp the OM to the string, the wall thickness
at the equator was increased to 13 mm, forming a flange.
%The  smoothness of the outer surface appears to be
%worse than that of the BENTHOS or VITROVEX spheres. This
%effect is negligable for the light collection since
%the refraction indices of water and glass are close
%to each other. However, a rippled surface
%looking upward in natural water is a good setting ground for
%sediments and biological matter. Thus, the sensitivity
%of upward looking OMs  may  detoriate with time (see below).
The hermetization of the OM along the equator is achieved 
by evacuating it via a special valve to 0.5\,-\,0.7 atm and sealing 
it by  homogenizing adhesive tape.

\subsection{Optical Contact}
The immersion material filling the gap between the bulb of
the phototube and the pressure housing should have 
{\it a)} a good transparency,
{\it b)} an index of refraction close to that
of glass and
{\it c)} high elasticity in order
to protect the bulb from deformation of the pressure housing
($\Delta D \approx$ 0.5 mm at 140 atm).

\vspace{1cm}

\begin{figure}[H]
\centering
%  \mbox{\epsfig{file=borglas.eps,height=9cm, angle=-90}}
\mbox{\epsfig{file=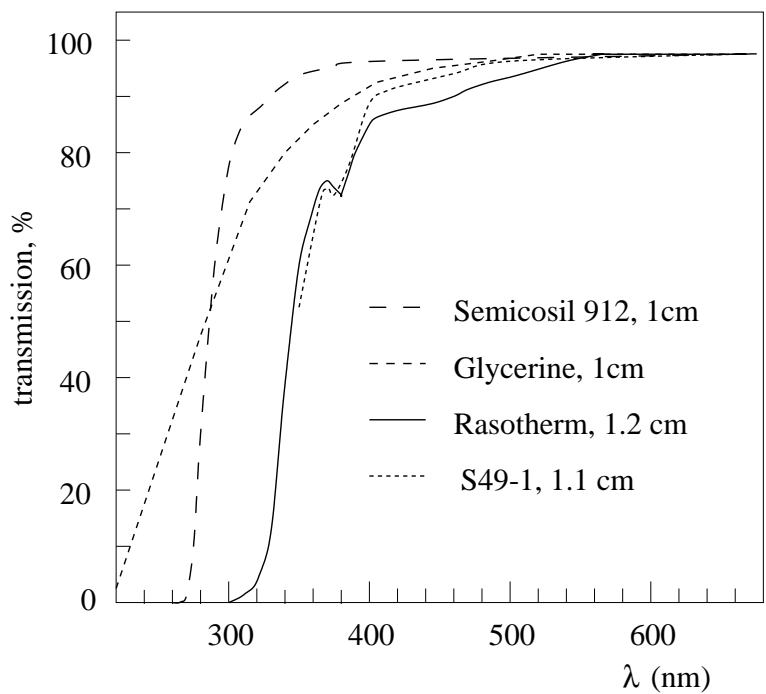,height=9cm}}
\caption[4]{\small
 Transmission curves of the EKRAN spheres (S49-1),
the VITROVEX spheres (Rasotherm), 1.1 cm glycerine and
 1.2 cm SEMICOSIL jelley. All curves are normalized to
their maximum transparency at high wavelength.
}
\end{figure}

In the standard approach, optically transparent silicon jelleys
are used \cite{Mat89,EOM,PRC}.
We have developed an alternative, new method:
The gap between tube and housing is filled with glyzerine
and sealed by casting a liquid polyurethane
compound  to the glyzerine surface.
The compound, being lighter than glyzerine, polymerizes
and forms a stable sleeve. The sleeve prevents the glyzerine
to leak into the  back hemisphere of the OM which houses
the HV supplies and other electronic components. It
fixes the position of the tube and at the same time does
not prevent the minor displacements necessary to balance
the pressure deformations of the housing.  
The advantages of this method are the following:
{\it a)} the index of refraction of glyzerine practically 
coincides with that of glass ($n$ = 1.47), 
{\it b)} there is no "delamination" of the immersion material
from the glass, a phenomenon
easily appearing when working with jelley, {\it c)}
the cost is low. The disadvantage is the risk that the
polyurethane might leak in which case not only the 
optical contact is lost
but also the glyzerine may corrode the electronics.

In parallel, we use the standard method which we tested
first in 1992. The gap
is filled with a two-component silicon jelley (SEMICOSIL, produced
by WACKER, Germany) with an index of refraction $n \approx$ 1.40.
Eight OMs (VITROVEX spheres) with SEMICOSIL jelley were
underwater for one year in 1992/93,
without showing any degradation of optical or
mechanical characteristics. Presently, we use
VITROVEX spheres and SEMICOSIL jelley for about 10 \% of all OMs.

The transparencies of pressure housings, glyzerine and jelley 
are shown in \linebreak Fig. 13 as a function of wavelength.

\subsection{Hermetic connectors}
The design of our connectors and penetrators started from 
the vacuum-proofed HF connector SRG-50-863
produced 
%by the OKTJABR company in Kamensk-Uralsk, 
in Russia.
The connector has an impedance of 50 Ohms and withstands  
working voltages up, to 500 V and temperature extremes
from -50$^o$C to +155$^o$C.
Following the experience we had gained formerly with connectors
produced by SEACON (USA), we modified
the SRG-50-863 for deep underwater applications. 
%In contrast to SEACON, the bulk head contains a glass insulator,
%soldered to the metallic body.
%The insulator is stable and 
The new connector is hermetic up to a pressure
of 200 atm. 
%Another difference is the electrical contact between
The outer screen is in electrical contact with water. 
In salt water this would result in strong electro-corrosion;
in fresh water, however, it is of negligible relevance.
The hermetic connectors and penetrators developed
in cooperation with the AKIN laboratory, Moscow,
can be operated
at all depths of Lake Baikal, i.e. down to 1.7 km. 

%Fig. 14 sketches
%the construction of the connector.
%The male part is mounted to a HF
%cable with a rubber jacketing. Since this cable
%is not designed for underwater applications, we 
%deposited an additional 1 mm jacket to the original cable.
%The diameter of the final cable is 0.6 cm.

\section{Operational Principle}

\subsection{Electronics}
The electronics, the trigger formation
and the data aquisition system of the {\it NT-200} Telescope
have been sketched in \cite{APP}. Here, we describe in more detail
the front-end electronics which is closely connected to the
operational principle of an OM pair. It is housed in the OMs 
itself as well as in the Svjaska Electronics Modules (see Fig. 1). 
Fig. 14 shows a block diagram of the components.

{\it a) Optical Module}

The OM houses two DC-DC HV supplies, one with a fixed output voltage
(presently 25 kV) for the {\it QUASAR} optical preamplifier, 
the other for the small PMT {\it UGON}, with a  voltage remotely controllable 
in steps of 10 V from 1.00 to 2.27 kV. Both supplies can be
remotely switched off/on. The anode signal is fed to an amplifier (10x),
the signal from the 11th dynode to an inverting 
amplifier (3x).
The amplifiers are mounted to a printed board. The
voltage divider for the {\it UGON} is integrated to
the photomultiplier itself. 
For amplitude calibrations, a LED is mounted close to the {\it QUASAR} 
photocathode. Its light level can be changed from 1 to 1000 
photoelectrons.

\bigskip

{\it b) Svjaska Electronics Module}

The anode signals from the two {\it QUASARs} are processed by 
the {\it local trigger board}. It consists of two 2-level
discriminators $D_1$ and $D_2$ as described in section 3, one for
each OM, and a coincidence circuit. The threshold 
of $D_1$ is set to 0.25 $a_1$, with $a_1$ being
the mean pulse height of a UGON 1-p.e. signal. The 
threshold of $D_2$ is remotely adjustable in the range 0.1 -- 10 $A_1$,
with $A_1$ corresponding to 1 photoelectron emitted from
the {\it QUASAR-370} photocathode. The output signal
from $D_1$ has to be confirmed by a signal from
$D_2$ (coincidence in Fig. 9). 
The output pulses from $D_1$ have 15 nsec length and are switched
in coincidence in a way, that the leading edge of  the output signal
({\it "local trigger"}) 
is determined by the first of the two input signals. 
In this way, late pulses are suppressed.

The dynode signals of a pair of OMs are led to the
{\it Q-T} module and summed by an 
analog summator. Each summator input can be inhibited remotely
(i.e. each OM can be excluded individually from the sum).
The sum signal is processed by a charge-to-time
converter based on the $Q$-$T$ circuit 1101PD1 (russian
analogue to MQT-200 from LeCroy). A local trigger would strobe
the 1101PD1, and the input charge is converted, with a maximum
conversion time of 70 $\mu$sec.  The  width of
the resulting signal corresponds to the charge of the dynode
signal, the leading edge is set by the leading edge
of the local trigger and defines the timing. The signal
is sent to a string electronics module one level higher
in the hierarchy and fed into TDCs
which digitize the time (11 bit) and the time-converted amplitude 
information (10\,bit) \cite{APP}.
 
The {\it Q-T} module can be operated in two modes. The first
uses a time conversion factor just as high that  a 
1-p.e. signal corresponds to 1 channel of the 
amplitude digitizing TDC. 
In the second ("calibration") mode, the conversion time 
is stretched and one photoelectron corresponds to 20 TDC bins.

\vspace{1.5cm}

\begin{figure}[H]
\centering
  \mbox{\epsfig{file=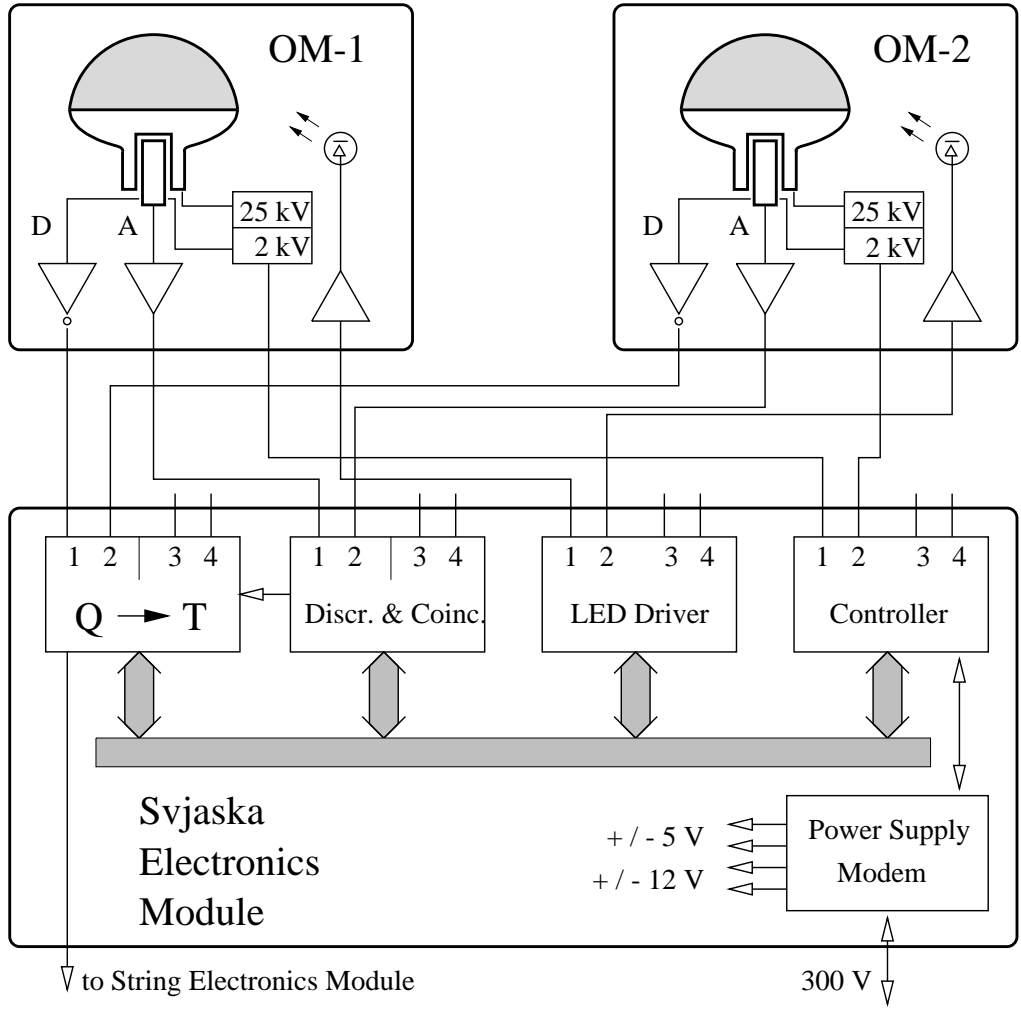,height=13.5cm,angle=-90}}
\vspace{0.8cm}
\caption[4]{\small
Scheme of the svjaska electronics controlling 2 pairs of
OMs (only the
2 PMTs belonging to the first pair are shown).
}
\end{figure}

\subsection{Single OM versus OM Pair}
In the course of the development the projects DUMAND and
BAIKAL, there have been long
 discussions about the advantages and drawbacks of
operating the PMTs as single detectors or as pairs. We
have been favouring the pair principle due to the following
reasons:

{\it Firstly}, the average counting rate per PMT is 
{\it in situ} (0.5-1)$\cdot 10^5$ Hz 
and, due to bioluminescence, seasonally reaches
 (2-3)$\cdot 10^5$ Hz. The coincidence reduces the rate 
to 100-300 Hz per pair typically. This low counting rate
is of significant advantage for the following goals:

\begin{itemize}

\item[a)] {\it data transmission and trigger formation}

The hard local coincidence allows to transmit all local 
signals to the underwater array trigger module just above
 the detector, to form an overall
trigger, to read out all signals and to transmit
digitized times and amplitudes via wire cables to shore.
Due to the low rate, a simple underwater hardware trigger
(like e.g. "$\ge$ 3 local triggers
in the whole array within 500 nsec") already gives a 
sample nearly free of accidental coincidences.

\item[b)] {\it track reconstruction}

In experiments operating the PMTs in single mode, background hits 
due to PMT noise, bioluminescence or K$^{40}$ are mixed into practically
every event \cite{Stenger}. These hits have to be eliminated by various
criteria and repeated fitting procedures rejecting those
PMTs with the highest time residuals. For the NT-200 detector,
the average number of hits {\it not} due to the muon track
is only 0.03/event, compared to about 10/event for an 
Ocean experiment operating $\approx$ 200
PMTs in single mode \cite{Stenger}. No coincidence between
distant PMTs reaches the noise hit rejection capabilities
of the local coincidence, due to the small
coincidence window of the latter.

\item[c)] {\it Search for slowly moving bright objects like magnetic
GUT  monopoles}

The detection principle is the registration of an excess in 
counting rates over time windows of the order of 10$^2$ $\mu$sec.
The rate excesses are buried in the
noise signals if the PMT is operated in single mode. Furthermore,
non-poissonian fluctuations of a single PMT might
fake a monopole event. Noise rates as well as non-poissonian effects
are effectively suppressed by the coincidence (see Fig. 15).

\end{itemize}

{\it Secondly}, "late pulses" are strongly suppressed.
These are pulses delayed by 10-100 nsec
due to (undetected) elastical backscattering of the photoelectrons
and multiplication after their second incidence on the
dynode system. Since it is rather unlikely that
both PMTs give a late pulse and since the response time is derived
from the first PMT yielding a signal, only for a very small
fraction of events the response time is that of a late pulse.

The time resolution is on the one hand worsened
since the azimuthal position of the PMT is unknown 
($\Delta x, \Delta y = \pm$ 30 cm, i.e. 1.5 nsec light travel 
time in water),
on the other hand, taking the time flag from the {\it first} 
hit PMT of a pair (convolution of eq. (4))
sharpens the time resolution. 
The two  effects almost balance each other.

At least for the {\it NT-200} project (given the
high external noise due to bioluminescence and the
robust "low-tech" philosophy of the electronics
and data transmission), these advantages prevail
the drawbacks which are:

\vspace{-2mm}

\begin{itemize}

\item[a)] A higher number of PMTs in order to instrument 
the same volume, 

\item[b)] a slightly higher threshold ($\ge$ 0.3 p.e. in 
{\it each} of the 
two PMTs compared to $\ge$ 0.3 p.e. in one PMT),

\item[c)] some mutual shadowing of the two OMs of a pair. 

\item[d)] possible signals in one PMT induced by the other PMT.
  
\end{itemize}

\vspace{1.0cm}

\begin{figure}[H]
\centering
  \mbox{\epsfig{file=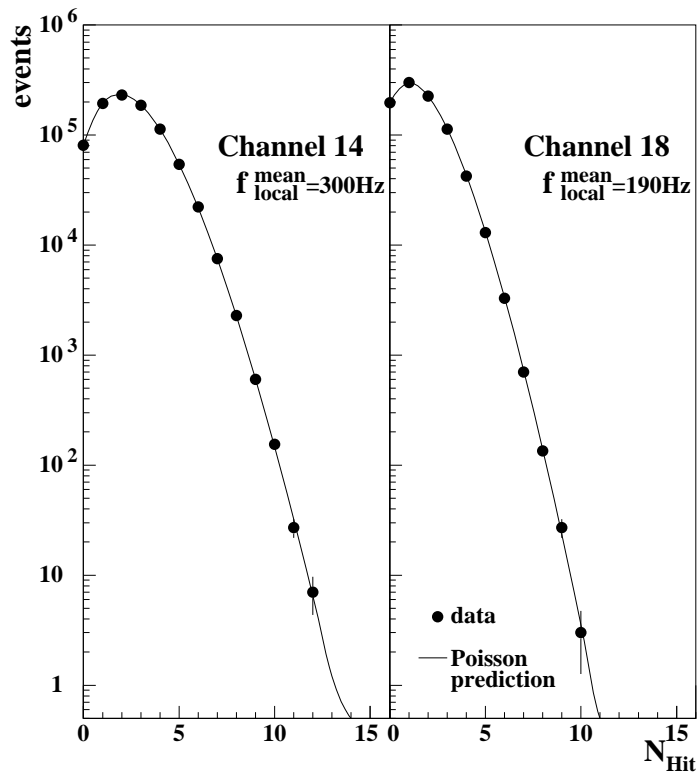,height=10cm}}
\caption[4]{\small
 Hit multiplicity distributions in a time window of 8 msec, as recorded
 with the monopole trigger system for two of the 18 channels of
  {\it NT-36}.
         Experimental data are indicated by points, the curve gives
         the Poisson prediction.
}
\end{figure}

\section{Calibration}

\subsection{In-situ tests}
In 1988/89, in-situ calibrations of cylindrical modules
containing a {\it QUASAR-300} (the early 30-cm variant of the
{\it QUASAR)} and a Philips {\it XP-2600} have been performed
in Lake Baikal. We used
a trigger telescope consisting of two tanks 
clamped to a string at a vertical distance of 6.5 m. The water
volume in the tanks was optically shielded from
the surrounding water, therefore the two flat-cathode PMTs watching
the tank interior were triggered only by Cherenkov
light from muons crossing the tank.  

A second string carried a pair of cylindrical OMs with the test
tubes.
The horizontal distance of this string with respect to
the string with the trigger telescope was varied
between 5 and 15 meters. Fig. 16 sketches the 
experimental arrangement and gives the registration
probability by the test tubes 
as a function of the distance between
muon and tubes. The curves are the results of
MC calculations based on the independently measured values
for water transparency, lensing effect and transparency
of the plexiglas
cap used at that time, and the photocathode sensitivity.

\bigskip

\begin{figure}[H]
\centering
  \mbox{\epsfig{file=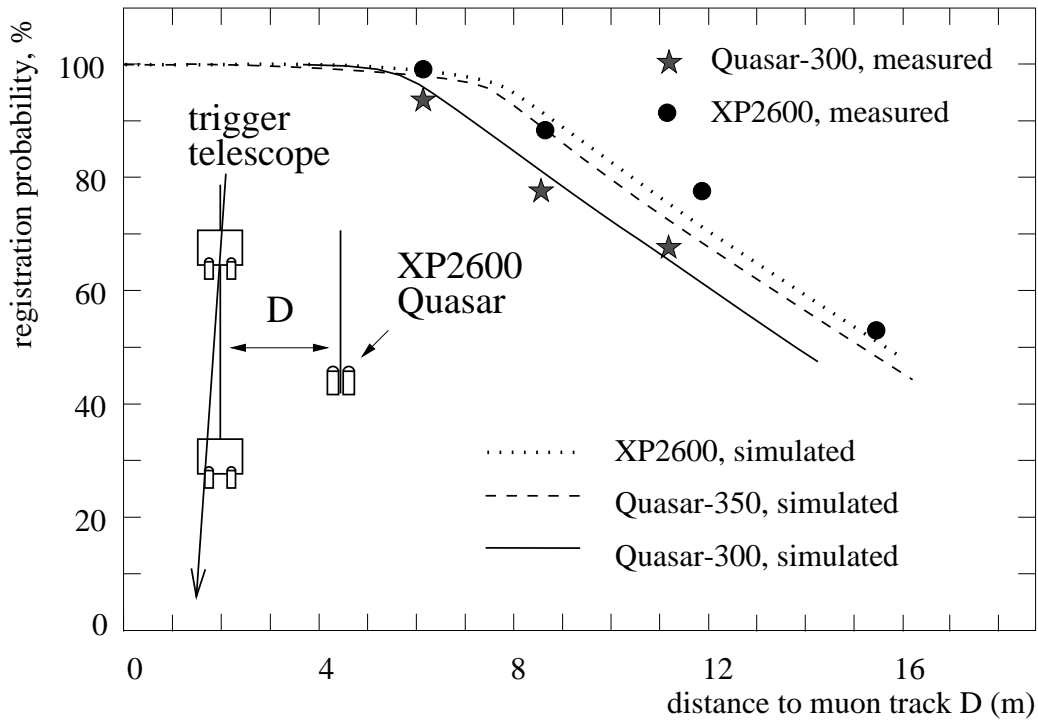,height=9.5cm}}
\caption[4]{\small
Determination of the registration probability of vertical
muons as a function of their passing distance to the OM.
{\it QUASAR-300} and {\it QUASAR-350} are the early versions of the
{\it QUASAR} tube, with 30 and 35 cm diameter, respectively.
}
\end{figure}

\subsection{Plane wave response}
A distant muon track illuminates an OM with a nearly plane wave
of photons. Given an incident flux of photons, $\Phi$ [photons/m$^2$],
the average number of photoelectrons is given by

\vspace{-2mm}

\begin{equation}
N_{PE} = \Phi \cdot F \cdot S(\theta).
\label{eq:5}
\end{equation}

Here, $\theta$ is the zenith angle with respect to the symmetry axis
of the OM, $S(\theta)$ the angular response normalized to unity
at $\theta$\,=\,0, and $F$ the absolute sensitivity at $\theta$\,=\,0.
$S(\theta)$ and $F$ include the relevant information needed for MC
calculations.

We have measured the response of OMs to a plane wave from a pulsed
LED and have determined $S(\theta)$ by rotating the OM in
the light beam. The experimental setup to measure the angular
dependence of the amplitude is shown in Fig. 17. The OM is mounted in
a black box  filled with water. It can be rotated an axis
perpendicular to the ligth front.
The OM  is illuminated with a green  LED
through a plexiglass window. The LED is at a distance of 2.5 meters,
the maximum deviation from planarity at the edge of the module is
4.3$^o$ for the box filled with water. The measured non-uniformity of
the light profile is less than 3\,$\%$ over the module cross section.

\begin{figure}[H]
\centering
  \mbox{\epsfig{file=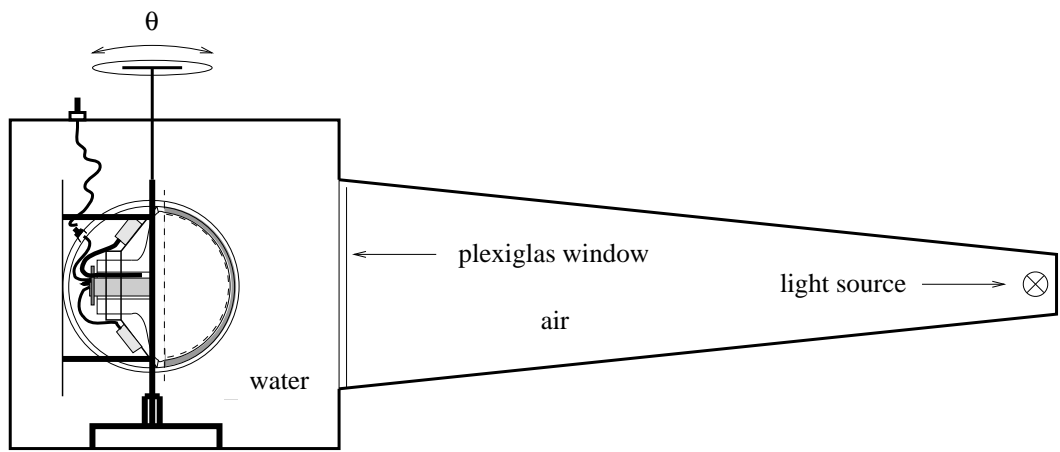,height=14cm, angle=-90}}
\caption[4]{\small
Plane wave test stand
}
\end{figure}

The results are shown in Fig. 18. Data points are normalized to the
signal at cos$\theta$ = 0. The deviation of the
curves from linearity is marginally. Neglecting the region at 
cos$\theta \ge$ 0.9, a linear fit 

\vspace{-2mm}

\begin{equation}
S(\theta) = A + B \cdot cos\theta
\label{eq:6}
\end{equation}

yields for the Baikal OMs $A = 0.49$, $B = 0.51$,
%\begin{center}
%\item $A = 0.49$, $B = 0.51$,
%\end{center}
similar to the DUMAND Japanese OMs \cite{Mat89},
the DUMAND European OMs measured with the same setup
\cite{EOM}, and the AMANDA OMs.
Also shown in Fig. 18 is the result of an analytical 
simulation including all effects of absorption, refraction 
and reflection \cite{Mohr}.

\vspace{0.3cm}

\begin{figure}[H]
\centering
  \mbox{\epsfig{file=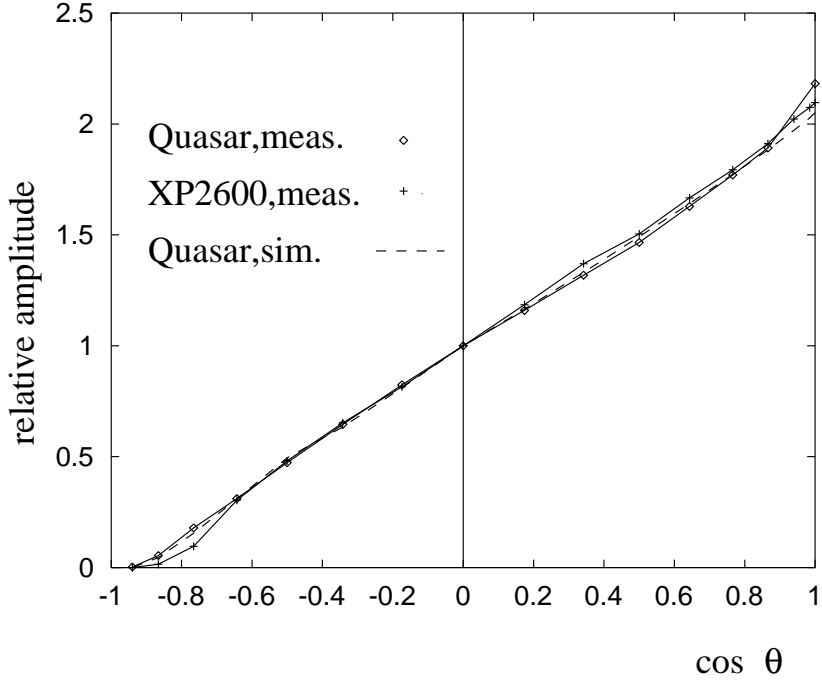,height=9.0cm}}
\caption[4]{\small
Angular sensitivity of the Baikal OM (labeled "Quasar") and
the Dumand European OM (labeled "XP-2600").
}
\end{figure}

\subsection{Amplitude and time calibration for the full telescope}
The data taking of the telescope is interrupted about twice
every week for calibration runs in order to determine the 
scale parameters of the amplitude and the time information.

{\it a) Amplitude}

The amplitude scale is calibrated by multi-photoelectron signals
from the LED. The average number of photoelectrons $N_{pe}$
of a charge distribution is derived from 

\begin{equation}
N_{pe} = A^2/D^2 \cdot (1 + d)^2
\label{eq:7}
\end{equation}

with $A$ and $D$ being mean value and dispersion of the
distribution, respectively,
and $d$ the relative dispersion of
a single-photoelectron signal. 
$N_{pe}/A$ gives the scale factor.
The high voltage 
for the {\it UGON} is changed
until one photoelectron corresponds to one amplitude channel.

The second (stretched) $Q$-$T$ mode allows to plot the 1-p.e. 
spectrum. This allows
an independent determination of the 1 p.e. scale factor
and measures also the  threshold value.

{\it b) Time}

The response time $t_i$ of an OM-pair $i$ with respect to
an arbitrarily choosen time $t_0$ is determined by two
calibration parameters,

\begin{equation}
 t_i = \beta_i \cdot n_i + \delta t_i
\label{eq:8}
\end{equation}

with $\beta_i$ being the scale factor for the time digitization,
$\delta t_i$ (in nsec) being the relative shifts of channel $i$
with respect to mean value of all channels, and $n_i$ the 
measured TDC-channel number for channel $i$.

In the calibration runs, the TDCs are started by noise
pulses of the PMTs and stopped by generator pulses with a
period $\tau$. From a distribution like the one shown in
Fig. 19, start and end point of the plateau,
$K_{min}$ and $K_{max}$, are determined with
an accuracy of 1\,-\,2 channels (1-2 nsec for a 10 bit TDC and 
$\tau = 1 \mu$sec). The $\beta_i$ are given by
$\tau / (K_{max} - K_{min})_i$, the flatness of the plateau
determines the differential linearity of the TDC, which is
better than 1 nsec in our case.

\begin{figure}[H]
\centering
  \mbox{\epsfig{file=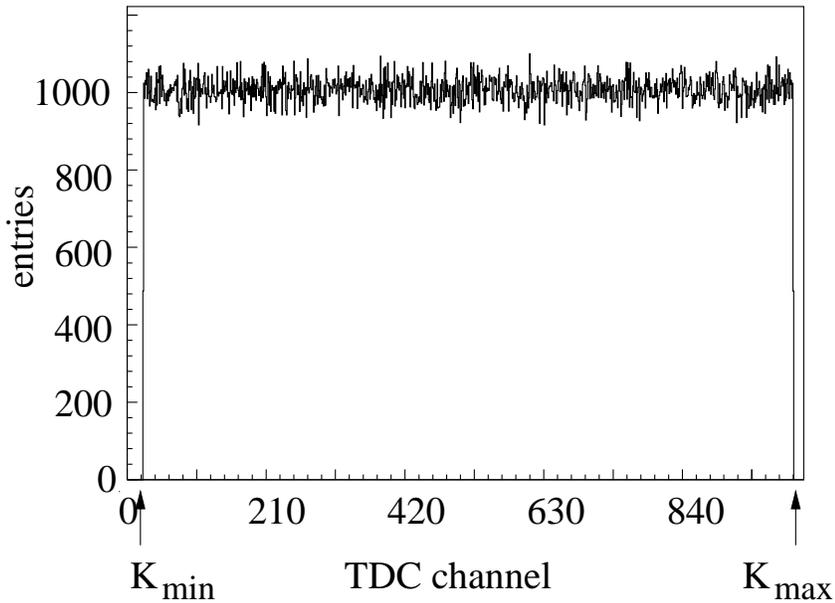,height=8cm}}
\caption[4]{\small
Determination of the scale factors $\beta_i$ for time
digitization (see text).
}
\end{figure}

The time shifts $\delta_i$ are determined with a help of
a calibration laser. This nitrogen laser, and a dye laser
shifting the original wavelength of 337 nm to
a spectrum peaking at 475 nm, are housed in a glas
cylinder just above the telescope. The light pulses of
less than 1 nsec width are guided via optical fibers 
of equal length to the OMs, with one fiber illuminating
one OM pair. A laser pulse generated a "laser" event
with typically most of the channels firing. The time shifts
$\delta_i$ are given by \cite{Thomas}:

\begin{equation}
\delta_i = \frac{1}{n_{ch}} \cdot 
\sum_{j=1}^{n_{ch}}{\frac{1}{n_{ji}}} 
\cdot \sum_{l=1}^{n_{ij}}{(\beta_i t_{il} - \beta_j t_{jl})}
\label{eq:8}
\end{equation}

with the first sum running over the total number
of channels, $n_{ch}$, and the second sum over all
events $n_{ij}$ with both channels $i$ and $j$ being
fired.
$\beta_i$, $\beta_k$ are the scale factors for channels
$i$ and $j$,  and $t_{il}$ and $t_{jl}$ are
the time codes of channel $i$ and $j$ within event $l$.

Fig. 20 shows the time difference distribution {\it after} the
correction procedure for the second and the sixth channel
of the first string of the {\it NT-36} array. The FWHM of
the laser peak is 2 nsec and the peak position is determined
with an accuray better than one nsec. The right peak
is due to downward going muons which had also triggered
the two channels separated by 25 m 
($ \approx 83$ nsec $\cdot$ 0.3 m/nsec).

\vspace{0.5cm}

\begin{figure}[H]
\centering
  \mbox{\epsfig{file=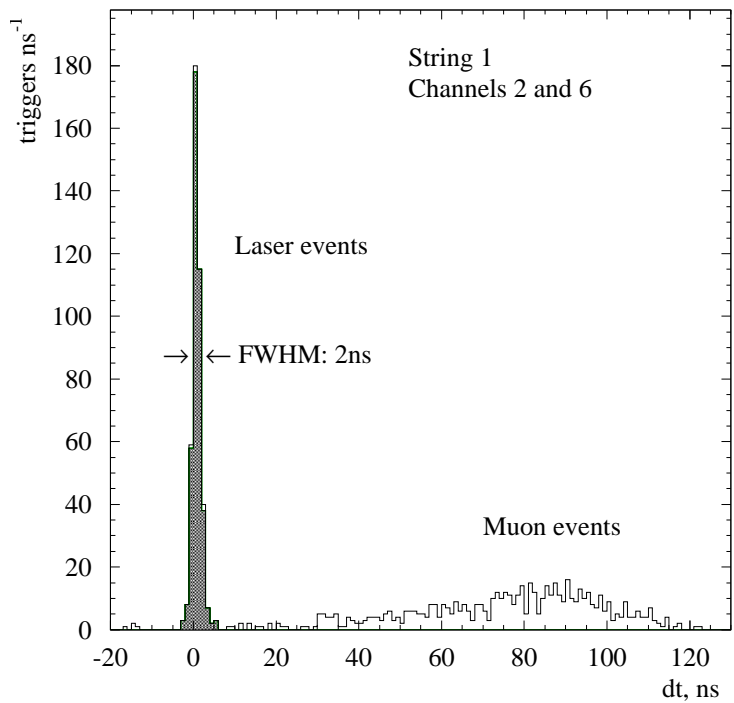,height=10cm}}
\caption[4]{\small
Distribution of light arrival time differences
         $\Delta t_{i,j,} = t_j - t_i$ for channels 2 and
         6 (string 1) for a typical laser run, with muon
         triggers recorded in parallel.
}
\end{figure}

We observed a drift in the $\delta$ values over
several months, presumably
due to changes in the speed of light in the fiber
under long-time pressure or water diffusion. These
small effects can be corrected {\it a posteriori} 
by reconstructing
muon tracks and requesting an average time residual
like that observed in MC calculations.
The overall accuracy of the time calibration is about 2 nsec.

\section{Long Term Operation Underwater}

\subsection{Reliability}
The first year of {\it NT-36} has demonstrated the stability and
reliability of all mechanical elements of the OMs. None of the OMs
did leak, none of the  polyurethane sealing layers
 and none of the QUASAR tubes have been damaged. 
Until 1996, nearly thousand penetrator/valve holes have been drilled
in more than 130 spheres (OMs and electrical modules) which afterwards
have been operated over one to four years at 1.1 km depth.
None of the feedthroughs did leak. Only 2 spheres -- in 1995 -- leaked
slightly due to cracks which had developed after a year underwater.
This effect was clearly due to a manufacturing error eliminated
in the mean time.

An unexpected problem was discovered with the power lines
submitting\linebreak 300 V to the electronics modules. In 4 out of
30 connectors a parasitic current between the central
wire and earth appeared, leading  to
strong electrolytic currents
across the water to the string and particularly to the
failure of the 1993 acoustic coordinate monitoring system
driven via one of these connectors. The reason for this effect seems
to be that, under the pressure of
water, the plastificator from the PVC jacket is squeezed into the
connector. This does not influence the functionality of the HF lines,
but obviously leads to the formation conducting channels in the
300\,V connectors. In the mean time, the jackets of the cables
have been improved an the effect has disappeared.

Another problem was the initially unacceptably high failure rate
of some electronic components. The percentage of working channels,
averaged over the full year, was only $S \approx$ 70\,\% 
in 1993/94 ({\it NT-36}).
(An array with a linearly decreasing number of living OMs,
starting with 100\,\% and ending after a year with 50\,\% living OMs,
would have $S$ = 75\,\%.) Losses where  dominantly due to failures
or misoperation of the HV supplies, secondly to failures of
the $SEM$ controllers. With $S$ still only 75\,\% in 1994/95, the
year 1995 was used for a total re-design of the 25 kV supply. 
This, and changes at the 2 kV supply as well as at the controllers
led to $S$ = 85 \% for the 1996 array {\it NT-96}. The goal for
the next years will be to increase $S$ up to 90\,-\,95\,\%. For the
OMs alone this number is already nearly reached, and next
improvements have to concentrate to other components of the 
detector.

In summary, the reliability optical module is suitable for long-term
underwater operation in a 200 OM array, taking into account that
yearly repair of failed components is possible. For arrays
larger by an order of magnitude, further significant improvements
are desirable.

\subsection{Sedimentation}

A phenomenon strongly influencing the sensitivity and, consequently,
the counting rates of upward facing modules, is sedimentation
of biomatter and dust on the upper hemispheres of the modules.

Fig. 21 shows the trigger rates for two different conditions
over a period of 225 days, starting with April 13th, 1993.
Firstly, for the case that
at least 4 upward facing channels have been hit (upper graph),
secondly, for the condition that at least 4 downward facing channels
have been hit (lower graph). Only channels operating all 225 days
have been included. In the second case, one observes
a slight decrease of the rates down to 85-90\% of its original value.
In contrast, the rate for the {\it upward} trigger falls down by
nearly
an order of magnitude.

\vspace*{1cm}

\begin{figure}[H]
\centering
  \mbox{\epsfig{file=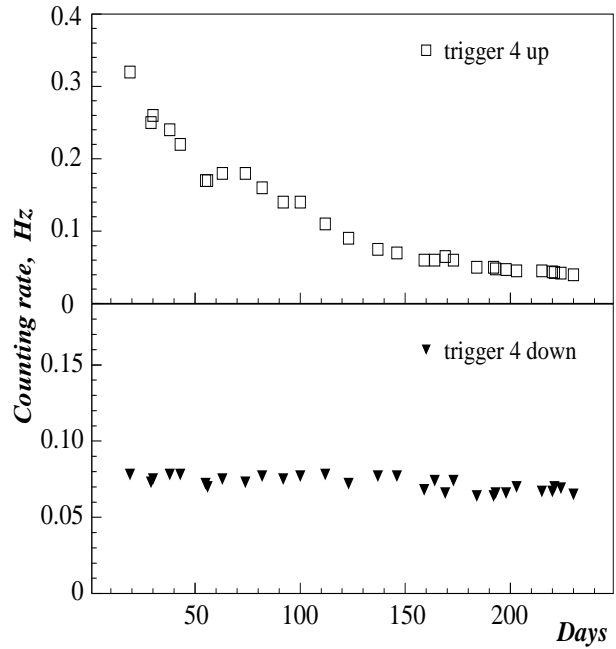,height=8.5cm,width=8cm}}
\caption[4]{\small
Trigger rates over a period of 225 days starting with
           April 13th, 1993. Top: Trigger {\it 4-up}
           (at least 4 upward
           channels hit). Bottom: Trigger {\it 4-down}
           (at least 4 downward channels hit).
}
\end{figure}

The inspection of the spheres after one year of operation showed that
sediments had formed a "hat" of bad transmission on the upward
facing hemispheres (see Fig. 22). The region near the equator was almost
free of sediments. This suggests to describe the variation of
the sensitivity $\eta$ of an optical module by the
following formula:

\begin{equation}
\eta = \eta_0 \cdot (p_1 + (1 - p_1) \cdot e^{-t/p_2})
\label{eq:9}
\end{equation}

where $t$ is the time after deployment in days. $p_1$ stands for
the part of the sensitivity contributed by the
equatorial region, the second
term describes the top region with exponentially decreasing
light transmission.

Replacing the sensitivity $\eta_0$ used in the Monte-Carlo
calculations by $\eta$ as defined above, and fitting the resulting
trigger rates to those experimentally measured in 1993 (1994), one
gets $p_1 = 0.33$ (0.36)  and $p_2 = 96.2$ (102.0) days
(numbers in brackets are for the 1994
array {\it NT-36$^{\prime}$}).
Consequently, the sensitivity of an upward facing module
to atmospheric muons decreases
to 35\,$\%$ after a year.
Both parameters change only slightly from year to year.
Note that the sensitivity of an upward facing OM
to upward going muons from neutrino interactions
is influenced less, since in average for these tracks the
equatorial part of the module is illuminated stronger
than the top region. Presently, we are looking for methods
to reduce sedimentation effects. E.g., the accumulation
of sediments can be reduced by
a smoother OM surface or by dressing the OM with
a plexiglas cone (see Fig. 22). 

\vspace*{0.4cm}
\begin{figure}[H]
\centering
  \mbox{\epsfig{file=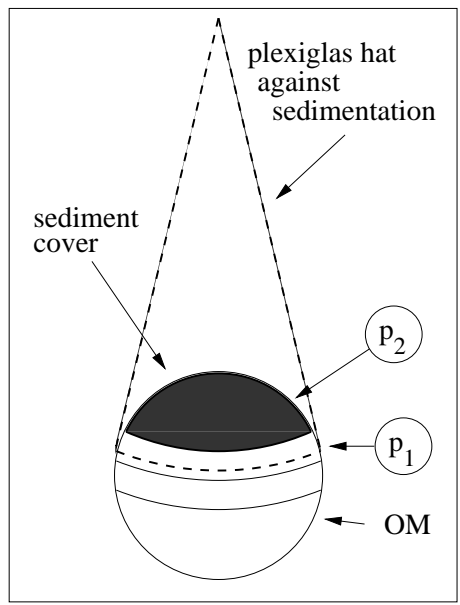,height=6cm,angle=-90}}
\vspace{0.4cm}
\caption[4]{\small
Sedimentation on upward facing modules (region $p_1$ --
negligable sedimentation, region $p_2$ -- exponentially decreasing
sensitivity due to sedimentation(see eq. 9). The dashed lines indicate
the
plexiglas cone to prevent sedimentation on the sphere.
}
\end{figure}

The straight-forward solution is of course to direct all OMs downward.
This increases the sensitivity to upward muons from neutrino 
interactions by\linebreak \mbox{20\,-\,30\, \%},
slightly reduces the identification capabilities with
respect to fake event (downward muons faking upward muons),
and limits the precision of downward muon physics.
In the presently operating
array {\it NT-200}, 160 of the 192  OMs face downward.

\subsection{Track Reconstruction}
During 4 years of data taking with the stepwise increasing
stages of the Baikal Neutrino Telescope we have accumulated
technical and methodical experience as well as 
first relevant results.  Physics results from {\it NT-36} are
reported in \cite{APP}, and from {\it NT-96}
in \cite{ICRC97,Erice}.

\begin{figure}[H]
\begin{minipage}[b]{6.7cm}
\hspace*{0.5cm}
\epsfig{file=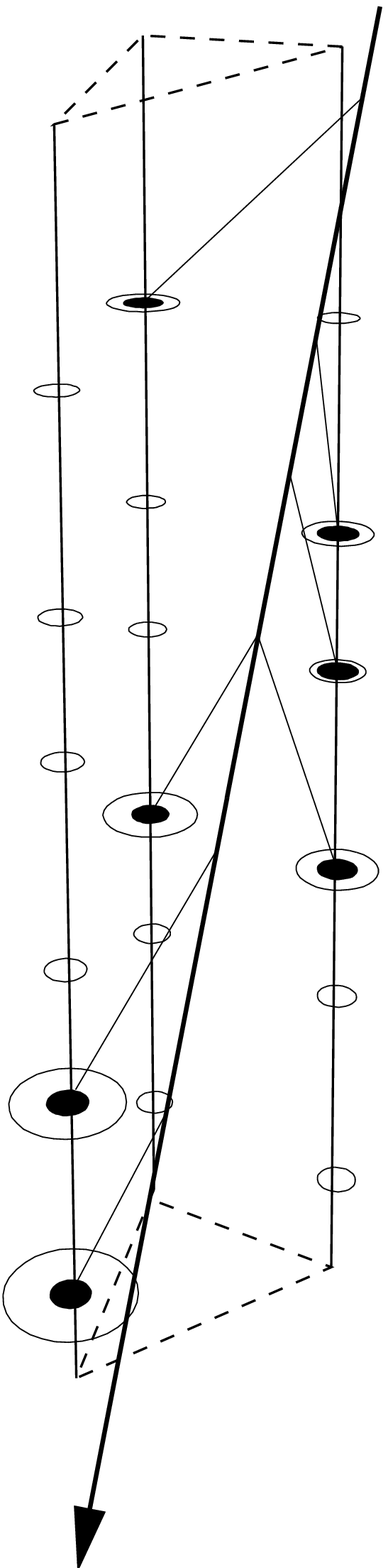,height=7.5cm,width=5cm}
  \caption [10]
  {Single muon event recorded with {\it NT-36}.
Hit channels are in black. The thick line gives the
reconstructed path, thin lines pointing to the channels mark
the path of Cherenkov photons as given by the fit to the
measured times.
The sizes of the ellipses are proportional to the
recorded amplitudes.
}
\end{minipage}
\hfill
\begin{minipage}[b]{6.7cm}
\epsfig{file=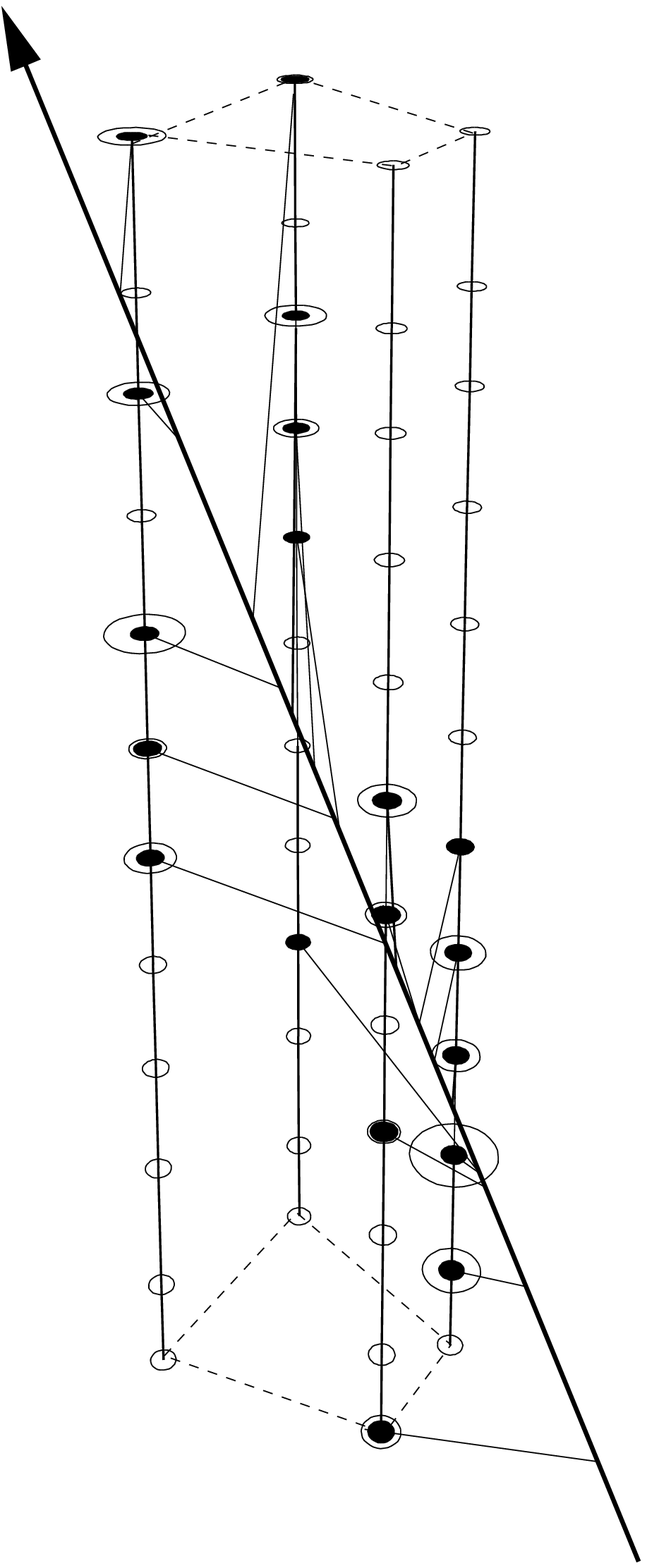,height=7.5cm,width=5cm}
\vspace{2.5cm}
          \caption [11]
           {
 A neutrino event with 19 hits, recorded with {\it NT-96}.
The fake probability of this event is
smaller than 1\,\%.
}
\end{minipage}
\end{figure}

The initial test an
underwater telescope has to undergo is the correct
reconstruction of atmospheric muons. They enter the
array from above and are recorded
with a frequency of several Hz. 
Fig. 24
shows a typical single muon  event
firing 7 of the 18 channels of {\it NT-36} and
reconstructed with a $\chi^2/NDF = 0.57$.
Monte Carlo calculations using as input the timing and amplitude properties
of the {\it QUASAR} measured in laboratory do well reproduce the
experimental
data like amplitudes, time differences and angular distribution \cite{APP}.

The crucial demonstration of the
functionality of a neutrino telescope
is the identification of the rare neutrino
events among the huge amount of atmospheric muons. Their
signature is given by a muon entering the array from {\it below}.
Still too small to detect the feable fluxes from extraterrestrial
neutrino sources, 
{\it NT-200} will be used to investigate those neutrinos which
have been generated in the atmosphere 
in interactions of charged cosmic rays ("atmospheric neutrinos")
and to search for neutrinos due to dark matter annihilation
in the center of the Earth \cite{Erice}.
In {\it NT-200}, about one neutrino per day will be recorded.
With {\it NT-36} and {\it NT-96}, 14 neutrino events have been
identified, in accordance with Monte Carlo estimates. 
Fig. 25 shows one "gold plated" event with 19 hits.

%\bigskip

%{\large \it  Limnology}

%Apart from its function as a neutrino telescope, the Baikal detector
%can be used to monitor  water
%parameters. The array
%permanently records photomultiplier counting rates,
%and periodically parameters
%like optical transmission at various wavelengths, temperature,
%conductivity, pressure (CDT sondes), and speed of sound.
%These measurements form a unique data set
%which can be related to CDT measurements at other locations
%in order to build a comprehensive picture of water exchange
%processes in the lake.

%Counting rates of the PMTs
%are dominated by  water luminescence.
%Fig. 26 shows the  counting rate variations
%recorded with  {\it NT-36} in 1993/94
%and compares it to the bacteria concentration measured at
% a distance of 50 km to the {\it NT-200} site,
%at 10 m depth below surface.
%In August/September
%we observe an increase of the luminosity to extremely high levels.
%The changes of the local trigger rate are not
%reflected in the muon trigger rate, since the muon trigger is
%essentially dominated by atmospheric muons, with negligible
%contribution by random hits (water luminescence or dark noise).

%\begin{figure}[H]
%\centering
%  \mbox{\epsfig{file=bact.eps,height=13cm,angle=-90}}
%\caption[2]{\small
%Average counting rate of OMs vs. time, compared to
%bacteria concentration at surface.
%}
%\end{figure}

\section{Summary and Outlook}

We have constructed a deep underwater Optical Module (OM) 
which is the key component of the neutrino telescope in 
Lake Baikal. Most parts of the OM, like the phototube {\it QUASAR-370}, 
the pressure sphere protecting the {\it QUASAR}, electronics
as well as connectors and feedthroughs have been developed
by our collaboration, in close cooperation with industry.

Since 1993, we have been permanently operating configurations 
with a growing number of OMs. A number of scientifically relevant
results have been obtained, ranging from counting rate variations
which reflect water transport processes to a precise measurement
of the angular spectrum of atmospheric muons. Most
notably, first neutrino events have been unambiguously
identified. During all years of data taking, none of the
pressure housing did leak, and the reliability of the OMs has
been improved continuously.

The 144-OM array operated in 1997 was upgraded
to 192 OMs in April 1998. With this upgrade ({\it NT-200}), the
short term goal of the collaboration is completed.

The technical
solutions used until now turned out to be adequat for a first 
generation neutrino telescope like {\it NT-200}. In the next step, we 
envisage a telescope consisting of 1000\,-\,2000 OMs. For this
telescope, the OM design will change in various respects.
Firstly, an array much larger than {\it NT-200} calls for
higher reliability. Some basic peculiarities
of the Baikal telescope may change. The present electronics
was conceptually developed in 1988\,-\,1990. Clearly,
a new design of the electroncis is necessary, using
higher integrated circuitry, eventually omitting the "svjaska
electronics modules" and moving the corresponding functions
to the OM or to the string electronics modules, making
use of more advanced signal transmission techniques etc.
Almost for sure, a future OM will have less feed-throughs in order 
to further minimize the danger of leaks. This, in turn,
request changes in the electronics. For instance, one might
be forced to omit the separate dynode read-out for amplitude
mesurement and pay for that with a somewhat smaller dynamic
range.

As an example for future develoments of the OM  we focus here
on its main component, the {\it QUASAR}-tube.

Amplitude and  timing characteristics  of
the {\it QUASAR-370}  depend strongly on the ratio $\frac{G}{\tau}$
(eq. 3).
We  are investigating some
new scintillator materials like ScBO$_{3}$:Ce (SBO),
         YAlO$_{3}$:Ce (YAO) and  YClO$_{5}$:Ce (YCO), with
         2\,-\,3 times larger  light  yield  than
         Y$_{2}$SiO$_{5}$:Ce (YSO) and 28\,-\,30 ns decay time.
We have manufactured each 5 pilot tubes with SBO and YAO.
The average gain values of the corresponding preamplifier 
tubes turned out to be twice as high as for fo YSO,
$G$ = 50\,-\,60. The single photoelectron resolution is
less than 50\,\% and the time resolution about 1 nsec.

Technological improvements
like  the chemical protection of single scintillator grains are
expected to give a larger gain $G$ in  tubes
with ordinary Y$_{2}$SiO$_{5}$:Ce.
Other lines of improvement are directed to increase significantly the
photocathode sensitivity and to decrease the dark current rate. For
the present  version, the average dark current rate is
about 35 kHz at room temperature and half of that at 0$^o$C.

Another line of principial re-design and improvement is the
replacement of the scintillator screen and the small PMT by
an diode array or by a foil first dynode followed by
a mesh dynode chain. All these improvement make the {\it QUASAR}
increasingly interesting to other fields 
apart from underwater telescopes,
in particular for Air-Cherenkov Telescopes.

As mentioned in section 5.2, we prefered a pairwise operation of
OMs by different reasons, most notably effective suppression of noise
counts and elimination of prepulses, late pulse and afterpulses.
However, for much larger projects the high cost of this
approach may be not acceptable. In order to maintain the local
coincidence and, at the same time, to reduce the number of OMs,
we have developed a two-channel version of the {\it QUASAR}. We made use 
of the fact, that photoelectrons of the central part of the
{\it QUASAR} are focussed onto the central spot of the luminescent
screen, whereas photoelectrons from peripherical regions are collected
onto the circular edge area of the screen. We have developed  2 pilot
samples of a two-channel, small
(3cm diameter) version of the UGON with a mesh dynode system,
one channel collecting photoelectrons from the central region,
the other from the periphery. The cross-talks measured are
about 2\,\%. Switching the two channels in coincidence
results in a noise rate as low as 100 Hz.

\vspace{0.5cm}

\section {Acknowledgments}

We are indebted to G.N. Dudkin, V.Yu. Egorov, O.I. Gress, A.I. Klimov,\linebreak
G.L. Kharamanian, A.A. Lukanin, P. Mohrmann, A.I. Panfilov, V.A. Poleshuk,
V.A. Primin, I.A. Sokalski, Ch. Wiebusch and M.S. Zakharov for help
at various stages of development and tests of the optical module.


\newpage


\begin{thebibliography}{99}

\bibitem{APP} I.A. Belolaptikov et al.,  
%The Baikal Underwater Cherenkov Telescope: 
%Design, Performance and First Results, 
{\em Astroparticle Physics} {\bf 7} (1997) 263.

\bibitem{Proposal} I. Sokalski and Ch. Spiering (ed.) {\em The Baikal 
                   Neutrino Telescope NT-200 -- Project Description.} 
                   BAIKAL {\bf 92-03}.


\bibitem{Physics} see, for a review of physics missions: 
                  V.S. Berezinsky, {\em Nucl. Phys.} (Proc. Suppl.)
                  {\bf B31} (1993); 
                  T.K. Gaisser, F. Halzen and T. Stanev, 
                  {\em Phys. Rep.} {\bf 258} (1995) 173.

\bibitem{NT-36} first results from NT-36 have been published in
                I.A. Belolaptikov et al., {\em Nucl. Phys.}
                (Proc.Suppl.) {\bf 35} (1994), pages 290 and 301.  


\bibitem{XP-1} G. van~Aller,  S.O. Flyckt,  W. Kuhl,  
%An  electro-optical preamplifier combination with integrated power  
%supply offering excellent single electron resolution for DUMAND. 
{\em IEEE Trans. on Nucl. Sci.} {\bf NS-30} (1983) 469.


\bibitem{XP-2}  G. van~Aller et al., 
%A "Smart" 15  inch PMT, 
{\em Helvetica Physica Acta} {\bf 59} (1986) 1119.

\bibitem{Lub92} R.I. Bagduev et al., {\em Proc. Int. Conf. "Trends in 
                Astroparticle Physics"}, Teubner Stuttgart/Leipzig
                (1994) 132.

\bibitem{Lub93} L.B. Bezrukov et al., {\em Proc. 3rd NESTOR Workshop},
                Pylos 1993, Univ. Athens 1994, 645.

\bibitem{Q-300}    L.B. Bezrukov et al.,
%         Properties and test results of a  photon  detector  based  on
%         the combination  of   electro-optical   preamplifier   and
%         a small photomultiplier.
{\em Proc. 2nd  Intern. Symp. "Underground
Physics  87"},  Baksan  Valley (Moscow: Nauka 1988) 230.

\bibitem{Bajar} B. Lubsandorzhiev, PhD-Thesis, Moscow 1993 (in russian).

\bibitem{Dor93} A.A. Doroshenko et al., {\em J. Strength of
  Materials}, (1993) 61 (in russian).

\bibitem{Mat89} S.Matsuno et al., {\em Nucl. Instr. and Methods} 
                {\bf A276} (1989) 359.

\bibitem{EOM} P.C. Bosetti, {\em Proc. 23rd ICRC} (Calgary) vol {\bf 4}, 534,
              and U. Berson, Ch. Wiebusch,
              {\em Dumand Internal Report} DIR {\bf 5-93}, Aachen 1993.

\bibitem{PRC} A. Biron et al., Upgrade of Amanda-B towards Amanda-II,
{\em DESY PRC} {\bf 97/05}, and http://sgi.ifh.de/$\sim$csspier/proposal.html

\bibitem{Stenger} V. Stenger, {\em Proc. 2nd NESTOR Int. Workshop} (ed.
L. Resvanis), Athens 1992, 79.

\bibitem{Mohr} P. Mohrmann, Diploma Thesis, Berlin 1995 (in german).

\bibitem{Thomas} Th. Mikolajski, PhD Thesis, Berlin 1995 (in german).

\bibitem{ICRC97} S. Barwick et al., {\em Proc. 25th ICRC} (Durban
1997) vol {\bf 7}, 1.

\bibitem{Erice} Ch. Spiering et al., {\em Proc. Int. School of Nuclear
  Physics, Erice}, Sept. 1997, astro-ph/9801044.

\end{thebibliography}
\end{document}